\documentclass{aa}  
\usepackage{graphicx}
\usepackage{txfonts}
\begin{document} 

   \title{Hot magnetic halo of NGC\,628 (M\,74)
   \thanks{Based on observations obtained with {\sl XMM-Newton}, an ESA science mission with instruments and contributions directly funded by ESA Member States and NASA}
    }
\author{M. We\.zgowiec\inst{1}
\and R. Beck\inst{2}
\and M. Hanasz\inst{3}
\and M. Soida\inst{1}
\and M. Ehle\inst{4}
\and R.-J. Dettmar\inst{5}
\and M. Urbanik\inst{1}}
\institute{
Obserwatorium Astronomiczne Uniwersytetu Jagiello\'nskiego, ul. Orla 171, 30-244 Krak\'ow, Poland, \\
\email{markmet@oa.uj.edu.pl}
\and Max-Planck-Institut f\"ur Radioastronomie, Auf dem H\"ugel 69, 53121 Bonn, Germany
\and Instytut Astronomii, Wydzia{\l} Fizyki, Astronomii i Informatyki Stosowanej, Uniwersytet Miko{\l}aja Kopernika, ul. Grudzi{\k a}dzka 5/7, 87-100, Toru\'n, Poland
\and European Space Agency, European Space Astronomy Centre (ESA/ESAC), Camino Bajo del Castillo s/n, 28692 Villanueva de la Ca\~nada, Madrid, Spain
\and Ruhr University Bochum, Faculty of Physics and Astronomy, Astronomical Institute, 44780 Bochum, Germany}
\offprints{M. We\.zgowiec}
\date{Received; accepted date}

\titlerunning{Hot magnetic halo of NGC\,628}
\authorrunning{M. We\.zgowiec et al.}

 
  \abstract
   {
In several spiral galaxies that are observed face-on, large-scale ordered magnetic fields (the so-called magnetic arms) were found. 
One of the explanations was the action of the magnetic reconnection, which leads to a higher ordering of the magnetic fields. 
Because it simultaneously converts the energy of the magnetic fields into thermal energy of the surroundings, 
magnetic reconnection has been considered as a heating mechanism of the interstellar medium for many years.
Until recently, no clear observational evidence for this phenomenon was found. 
   }
   {
   We search for possible signatures of gas heating by magnetic reconnection effects in the radio and X-ray data for the face-on spiral galaxy NGC\,628 (M\,74), 
which presents pronounced magnetic arms and evidence for vertical magnetic fields. 
   }
   {
The strengths and energy densities of the magnetic field in the spiral and magnetic arms were derived, as were the temperatures and thermal energy 
densities of the hot gas, for the disk and halo emission.
   }
   {
In the regions of magnetic arms, higher order and lower energy density of the magnetic field is found than in the stellar spiral arms. The global temperature of the 
hot gas is roughly constant throughout the disk.
   }
   {
The comparison of the findings with those obtained for the starburst galaxy M\,83 suggests that
magnetic reconnection heating may be present in the halo of NGC\,628. The joint analysis of the properties of the magnetic fields and the hot gas
in NGC\,628 also provided clues for possible tidal interaction with the companion galaxy.
   }


   \maketitle
%

\section{Introduction}
\label{recon}

The past decades of radio observations, including sensitive polarimetry at higher frequencies, have increasingly provided evidence that the dynamics of the interstellar medium (ISM) is controlled by the magnetic field \citep[see e.g. the review by][]{beck04}. 
Although radio emission, both thermal and non-thermal, is mainly associated with the star-forming galactic arms, 
polarised emission from the large-scale ordered magnetic fields in some galaxies was found between the spiral arms \citep[e.g.][]{beck07}.
To date, no widely accepted theory exists that would explain the origin of these magnetic arms \citep{beck15}. One of the proposed mechanisms 
is magnetic reconnection, which converts energy that is stored in the magnetic field into thermal energy of the surrounding gas \citep{hanasz98}.
Because reconnection occurs in tangled magnetic fields, the removal of this component would lead to a higher contribution from 
the ordered component to the total magnetic fields. 

To test this hypothesis, we analysed the radio and X-ray data for 
the grand-design spiral galaxy NGC\,6946 \citep{wezgowiec16}, in which magnetic arms are present \citep{beck07}. The comparison 
of the properties of the hot gas and the magnetic fields in the spiral arms and the inter-arm regions showed
a slight increase in the temperature of the hot gas (and the corresponding thermal energy) in the inter-arm regions that was accompanied by lower energy densities and by a higher order of the magnetic fields \citep{wezgowiec16}. To follow-up our analyses, we studied another grand-design spiral galaxy, M\,83, which also presents 
ordered magnetic fields between the stellar spiral arms \citep{frick16}. Although the effects that were found in NGC\,6946 could not 
be confirmed in M\,83 at a high level of significance, the deep X-ray data allowed us to obtain reliable information about the halo gas. Together with the properties of the magnetic fields derived from the radio data, this in turn provided clues that reconnection heating might 
also occur in the halo of M\,83 \citep{wezgowiec20}. The possibility that magnetic reconnection might occur in galactic haloes has been theoretically discussed 
already by \citet{kahn93}, and a number of subsequent papers considered it as a source of heating \citep[e.g.][]{zimmer96,tanuma03}.

In this paper we present an analysis of radio and X-ray data for the spiral galaxy NGC\,628 in search for signs of reconnection heating. 
This face-on galaxy shows both pronounced magnetic arms and substantial clues for the existence of vertical magnetic fields \citep{mulcahy17}.   

NGC\,628 (M74) is a large grand-design spiral galaxy that presents broad and extended stellar arms. 
Although it resides within a small group of galaxies \citep{auld06}, the studies by \citet{kamphuis92} suggest that NGC\,628 did not interact 
with its companions within the past 1\,Gyr. This allows NGC\,628 to be regarded as an isolated galaxy. Although a significant number of \ion{H}{i} 
holes were found in this galaxy \citep{bagetakos11}, which suggests a vivid star formation, the rate of this star formation remains at a moderate level 
of 0.68\,M$_{\sun}$\,yr$^{-1}$ \citep{kennicutt11}. This might be explained by the findings of \citet{marcum01}, who argued that NGC\,628 has 
experienced an increased star formation activity over the last 500\,Myr that now decreases rapidly. The basic physical properties 
of NGC\,628 are presented in Table~\ref{astrdat}.

\begin{table}[ht]
        \caption{\label{astrdat}Basic astronomical properties of NGC\,628}
\centering
                \begin{tabular}{lc}
\hline\hline
Morphological type \tablefootmark{a}& SAc       \\
Inclination        & 7\degr     \\
Diameter D$_{25}$  & 10\arcmin  \\
R.A.$_{2000}$      & 01$^{\rm h}$36$^{\rm m}$42$^{\rm s}$\\
Dec$_{2000}$       & +15\degr 47\arcmin 01\arcsec       \\
Distance\tablefootmark{b} & 7.3\,Mpc    \\
Column density $N_{\rm H}$\tablefootmark{c}& 4.73$\times$10$^{20}$\,cm$^{-2}$\\
\hline
\end{tabular}
\tablefoot{
All data except for type, distance, and column density are taken from HYPERLEDA database -- http://leda.univ-lyon1.fr -- see \citet{makarov14}.\\
\tablefoottext{a}{Taken from \citet{devaucouleurs91}.}
\tablefoottext{b}{\citet{karachentsev04}.}
\tablefoottext{c}{Weighted average value after the HI4PI survey; \citet{hi4pi}.}
}
\end{table}

The radio emission from NGC\,628 has been studied by \citet{mulcahy17}, who analysed sensitive 
single-dish observations obtained with the 100 m Effelsberg radio telescope, 
as well as deep S-band (3\,GHz) Karl G. Jansky Very Large Array (VLA) observations in the compact C and D configurations. 
They found little radio emission in the central region of the galaxy, which likely 
results from a low abundance of the massive O stars that eventually supply the cosmic-ray electrons 
to the interstellar medium. Consequently, the authors derived the typical strengths of the magnetic fields 
in NGC\,628 (of the order of 9\,$\mu$G). The observations also showed that NGC\,628 possesses pronounced magnetic arms between the stellar spiral arms of the galaxy, similar to those found in NGC\,6946 \citep{beck15}. A strongly polarised feature in the northern part of the disk was especially interesting. It is not associated with any of the star-forming regions. 
Furthermore, sensitive Faraday rotation data allowed detecting signs of possible Parker loops, rising from the eastern spiral arm 
above the disk and into the halo, as well as a significant feature in Faraday depth that coincides with one of the 
many \ion{H}{i} holes in NGC\,628, also suggesting that enhanced vertical magnetic fields are present in the halo of this galaxy.

NGC\,628 has been observed three times with the XMM-Newton X-ray space telescope \citep{jansen01}. 
The details of the observations are listed in Table~\ref{xdat}. 
The results of observations 0154350101 and 0154350201 
have been published in a number of papers, most of which focused on the populations 
of discrete sources \citep[e.g.][]{sonbas10,walton11,lin12,earnshaw19}, and only \citet{owen09} performed the analysis of the diffuse
X-ray emission from NGC\,628, along with several other face-on galaxies. 
This study, however, aimed at the global properties of the galaxies. In this paper, we add new observations, 
performed by the authors in January 2021, and take advantage of the increased sensitivity of the entire X-ray data for NGC\,628. 
Our findings are compared to the results obtained by \citet{owen09} below. 

We analyse new X-ray observations of NGC\,628 and focus on the diffuse emission from the hot gas. Its properties 
in different areas of the galaxy are then compared to the properties of the magnetic fields inferred from the radio data presented by \citet{mulcahy17}. 
The comparison is performed to search for signs of magnetic reconnection, such as those reported 
for NGC\,6946 \citep{wezgowiec16} and M\,83 \citep{wezgowiec20}. The spectral analysis of the hot gas can also give an insight 
into the properties of the halo gas, which might help to confirm the detection of the vertical magnetic fields \citep{mulcahy17}.

\section{Observations and data reduction}
\label{obsred}

\subsection{Radio observations}
\label{radiored}

The radio data at S band (3\,GHz) presented in this paper were already thoroughly analysed by \citet{mulcahy17}. 
For the purpose of this paper, we re-reduced these data following the calibrations performed by the authors, 
but focusing only on the data observed in the most compact D configuration, which provide high sensitivity to large-scale emission and 
ensure a resolution that matches that of the X-ray data best. The maps obtained in the cleaning process 
were restored with an elliptical beam of 26$\arcsec\times$22$\arcsec$ and were later smoothed with a Gaussian beam to the resolution of 30$\arcsec$.

\subsection{X-ray observations}
\label{xrayred}

To process all data from the observations listed in Table~\ref{xdat}, we used the SAS 19.0.0 package \citep{gabriel04} 
and its standard procedures that allowed us to produce event lists for all three detectors of the European Photon Imaging Camera (EPIC). 
From these lists, we excluded data 
that were acquired during periods of an increased level of the high-energy background radiation. This step is necessary to allow a reliable analysis of the diffuse X-ray emission. To identify these periods, we created light curves and selected for further analysis 
 only those observation times during which the level of the high-flaring background was below a standard 
thresholds of 0.35\,cts/s and 0.4\,cts/s for the EPIC-MOS \citep{turner01} and the EPIC-pn \citep{strueder01} cameras, respectively.

\begin{table}[ht]
        \caption{\label{xdat}Characteristics of the X-ray observations of NGC\,628}
\centering
\begin{tabular}{rccccc}
\hline\hline
Obs ID          &EPIC-pn&EPIC-pn    &total    & clean   \\
                &filter &obs. mode  &time [ks]&time [ks]\\
\hline
0154350101      & T  & FF &  34.3   &  28.9     \\
0154350201      & T  & FF &  23.0   &  23.0     \\
0864270101      & T  & FF & 107.7   &  34.5     \\
\hline
\end{tabular}
        \tablefoot{
        T - thin; FF - full frame.
                }
\end{table}

Next, we checked the filtered event lists for possible soft proton flare contamination using the method 
developed by \citet{deluca04} that is implemented in the script\footnote{https://www.cosmos.esa.int/web/xmm-newton/epic-scripts\#flare} provided by the authors.
The event lists of all three observations were found to be free from soft proton contamination.

For each of the filtered event lists, images and exposure maps (without vignetting correction) were produced. This was done with the help of the {\sc images} 
script\footnote{https://www.cosmos.esa.int/web/xmm-newton/images}. In the data selection needed to create the images, 
we used only events with FLAG=0 and PATTERN$\leq$4 (EPIC-pn) or FLAG=0 and PATTERN$\leq$12 (EPIC-MOS). 
This ensured best-quality data, which are required to analyse the diffuse X-ray emission.
The images were produced in four energy bands, that is, 0.2-1\,keV, 1-2\,keV, 2-4.5\,keV, and 4.5-12\,keV, which cover 
the entire energy range of the XMM-Newton EPIC cameras. This was needed to effectively detect the point sources 
within the galactic disk, whose emission was later excluded from the regions prepared for the spectral analysis. 
The two lower energy bands, that is, 0.2-1\,keV and 1-2\,keV, were also used to create a hardness ratio map that 
helps to trace a relative change in the amounts of cooler and hotter gas throughout the disk. 
These steps were performed for each of the three observations. Next, the corresponding images and exposure maps 
were combined into final EPIC images using the SAS task {\sc emosaic}. The combined images were adaptively smoothed 
with a maximum smoothing scale of 30$\arcsec$ and a signal-to-noise ratio of 30. 
To allow a precise creation of the hardness ratio map, the map in the medium energy band (1-2\,keV) was smoothed 
using the same scales as were calculated for the soft energy band (0.2-1\,keV) with the {\sc asmooth} task. 

The source detection described above was performed with the SAS meta-task {\sc edetect\_stack,} which simultaneously 
uses event lists from all observations to maximise the sensitivity of the conducted search. The detected sources were excluded 
from the spectral regions that were prepared to analyse the diffuse X-ray emission from the hot gas. The size of the removed area 
was selected individually for each source so that only the brightest emission was excluded. This approach was desirable because 
the exclusion within the area of the full point spread function (PSF) of the instrument could lead to an unnecessary removal 
of the already faint diffuse emission. To account for these smaller areas, a power-law component was added to the model fitted to the resulting spectrum. 
This component was still needed to account for any possible emission from unresolved point sources that also contributes to the X-ray sky background.

In the next step, spectra from all regions were extracted with the use of the same flag and pattern selection as when the images were created. 
For the extracted spectra, response matrices and effective area files were created. Because the sizes of all regions are 
much larger than the PSF, a detector map (image in detector coordinates) was also used. 

For the available observations of NGC\,628, a proper modelling of the background is very difficult. 
The significant angular size of the galaxy means that most of the sensitive part of the field of view is covered. 
\citet{owen09} used the local background, which also included the detector contribution. 
We note here that their work focused on the analysis of the global spectrum of
the entire emission from the galactic disk. In this case, the background measured in the annular region outside 
of the source region can be justified. In this work, however, we analyse individual regions within the
galactic disk, most of which show a very low surface brightness. 
Due to the limited sensitivity of the available observations and the resulting low surface brightness 
of the emission in the studied regions, the use of the blank-sky background \citep{carter07} 
led to its oversubtraction. This could be caused by increased background emission coming from either observations with 
lower column density than that in the direction NGC\,628 (Table~\ref{astrdat}) or from observations with a stronger detector contribution, likely those 
distant in time from the ones that we used. The low surface brightness of most of our regions suggests 
that the contribution from the detector background probably is much more significant.
Therefore, we decided to account for the detector background only, using the filter wheel closed (FWC) 
data\footnote{https://www.cosmos.esa.int/web/xmm-newton/filter-closed}, but only from those observations that were 
performed close in time to the ones that we use. 
The FWC event lists were then filtered in the same way as the source event lists, and the 
background spectra were extracted using the same regions as in the extraction of the source spectra.

In the final step, the corresponding spectral products, source and background spectra, response matrices, 
and effective area files were combined (SAS task {\sc epicspeccombine}). The obtained 
background-subtracted spectra were binned to reach 25 counts per energy bin, 
which proved to be effective in a wide range of sensitivities \citep{wezgowiec16,wezgowiec20}.
The spectra were then fitted using {\sc xspec12} \citep{arnaud96} and the models described in Sect.~\ref{spectra}.

For the overlays, we also used the UV image observed with the XMM-Newton Optical Monitor and the UVW1 filter 
from the most recent observations (ObsID 0864270101). 
The data were reduced with the standard pipeline procedure ({\sc sas} meta-task {\sc omichain}).

\section{Results}
\label{results}

The radio data that we show in this paper were already presented by \citet{mulcahy17}. Nevertheless, we used the D-configuration data only, which result in high sensitivity to extended structures, especially of faint polarised emission, which is crucial for our study. 
In Sect.~\ref{radio} we therefore briefly present our maps with special attention to the features revealed by the more sensitive larger beam. 
In Sects.~\ref{dist} and \ref{spectra} we present the maps of the X-ray emission and its spectral properties, respectively. 
For the overlays, we used the H$\alpha$ map of NGC\,628 from the {\em SIRTF} Nearby Galaxy Survey \citep[SINGS;][]{kennicutt03}.

\subsection{Radio emission}
\label{radio}

\begin{figure*}[ht]
\resizebox{0.5\hsize}{!}{\includegraphics[clip]{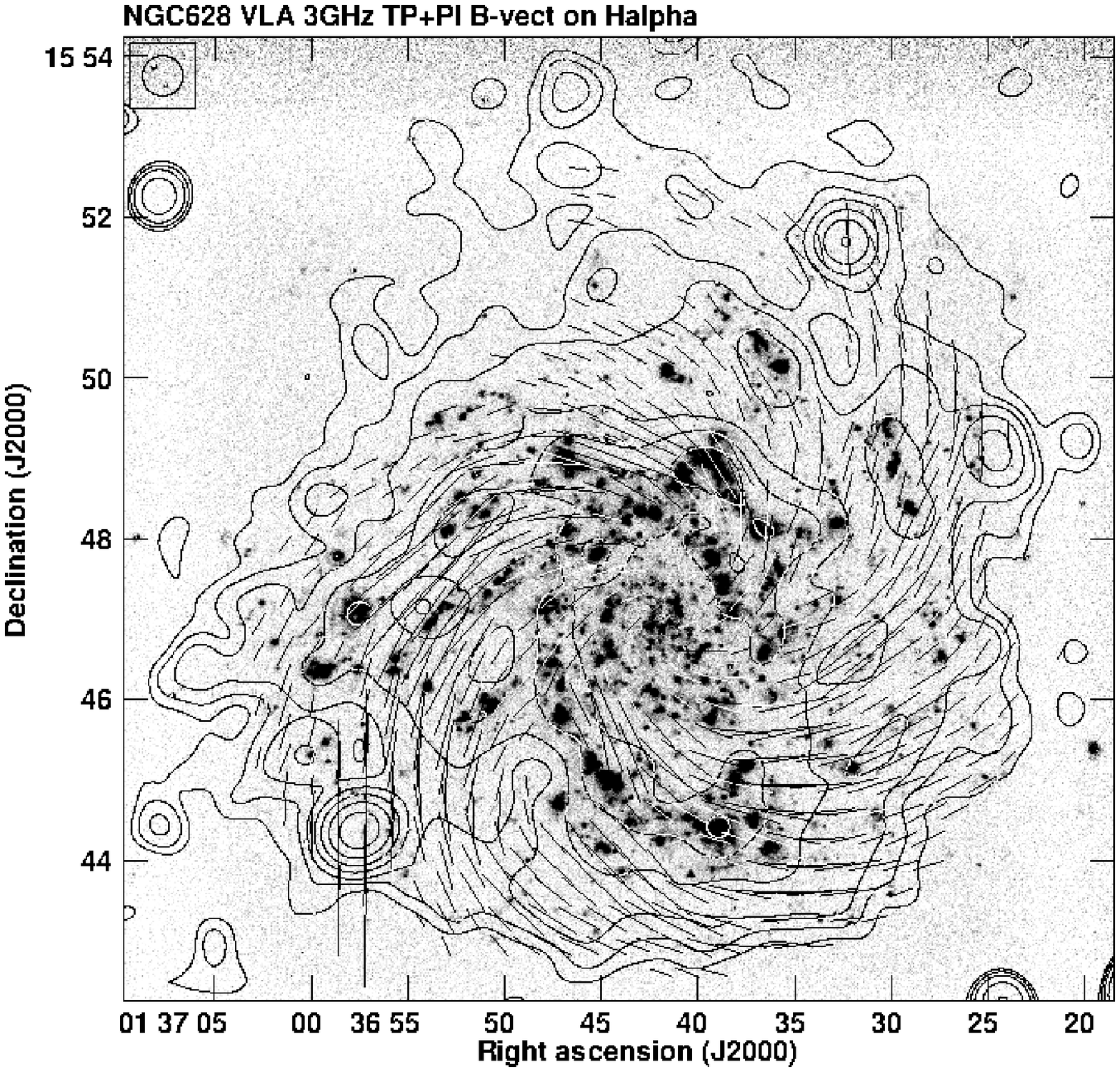}}
\resizebox{0.5\hsize}{!}{\includegraphics[clip]{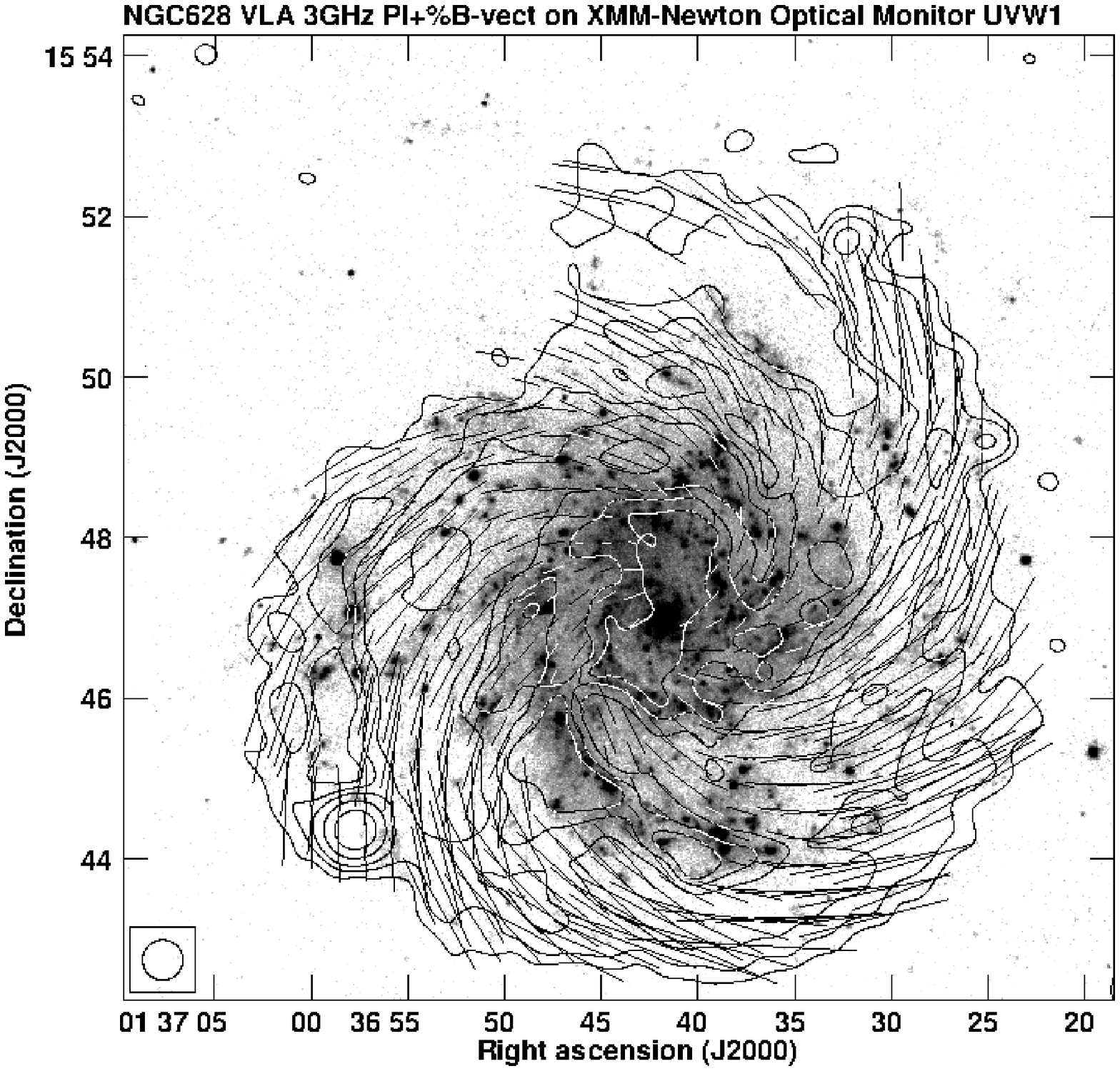}}
\caption{
        Map of the total ({\it left}) and polarised ({\it right}) radio intensity at 3\,GHz 
        ($\lambda$ 10\,cm) of NGC\,628 overlaid on an H$\alpha$ and XMM-Newton Optical Monitor in UVW1 filter map, respectively. 
        The contour levels are 3, 5, 8, 16, 32, 64, 128, 256, and 512 $\times$ 24\,$\mu$Jy ({\it left}) or 10\,$\mu$Jy ({\it right}). 
        The lines show the orientation of the magnetic fields, and their lengths of 1\arcmin are proportional to the polarised intensity 
        of 150\,$\mu$Jy ({\it left}) or to a degree of polarisation of 36\% ({\it right}). 
        The angular resolution is 30$\arcsec$.
        }
\label{10radio}
\end{figure*}

The map of the total radio intensity (Fig.~\ref{10radio}, left) shows a larger extent of the emission than of the star-forming disk, 
with a significant extension towards the north. The lines of the magnetic fields follow the spiral structure of the galaxy, 
although they seem to be phase-shifted and filling the inter-arm space. This is especially visible in the right panel of Fig.~\ref{10radio}, 
where the contours of the polarised radio emission are presented together with the magnetic field vectors with their lengths proportional 
to the degree of polarisation. The lower resolution of this map compared to that of \citet{mulcahy17} allowed us to reveal 
that the western magnetic arm extends far to the north of the galaxy and turns eastward. The larger beam also recovered more polarised 
emission in the central part of the galaxy, but it is still strongest in its outer parts. Nevertheless, the total and polarised radio fluxes 
agree within the uncertainties when C+D and D array maps are compared (Table~\ref{radiofluxes}).

\begin{table}[ht]
        \caption{\label{radiofluxes}S-band total radio flux densities (in mJy) of M\,74}
\centering
\begin{tabular}{rccccc}
\hline\hline
Map                & C+D array\tablefootmark{a} & D array\tablefootmark{b}   \\
\hline
Total intensity    & 120$\pm$6                  & 119$\pm$6  \\
Polarised intensity& 16$\pm$1                   & 17$\pm$1  \\
\hline
\end{tabular}
\tablefoot{
        \tablefoottext{a}{measured in the maps from \citet{mulcahy17}.}
        \tablefoottext{b}{this paper.}
}
\end{table}

Furthermore, east and north of the galactic centre, 
local maxima of the polarised emission are visible that do not correspond to any star-forming region. The northern maximum was already 
investigated by \citet{mulcahy17}, who found an exceptionally strong and ordered magnetic field in this position that may be compressed 
by an expanding super-bubble.

\subsection{Distribution of the X-ray emission}
\label{dist}

\begin{figure*}[ht]
        \resizebox{0.5\hsize}{!}{\includegraphics[clip]{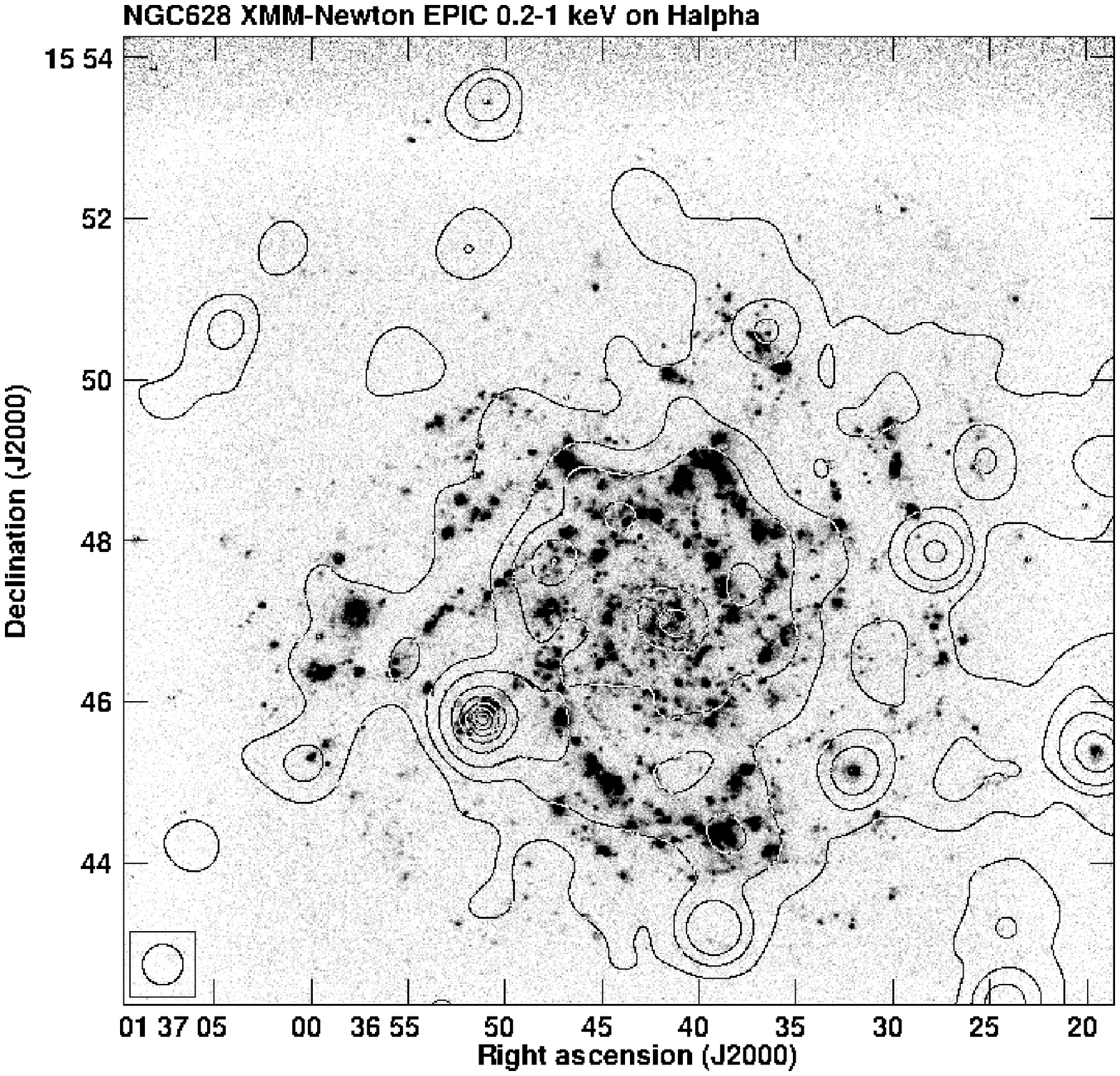}}
        \resizebox{0.5\hsize}{!}{\includegraphics[clip]{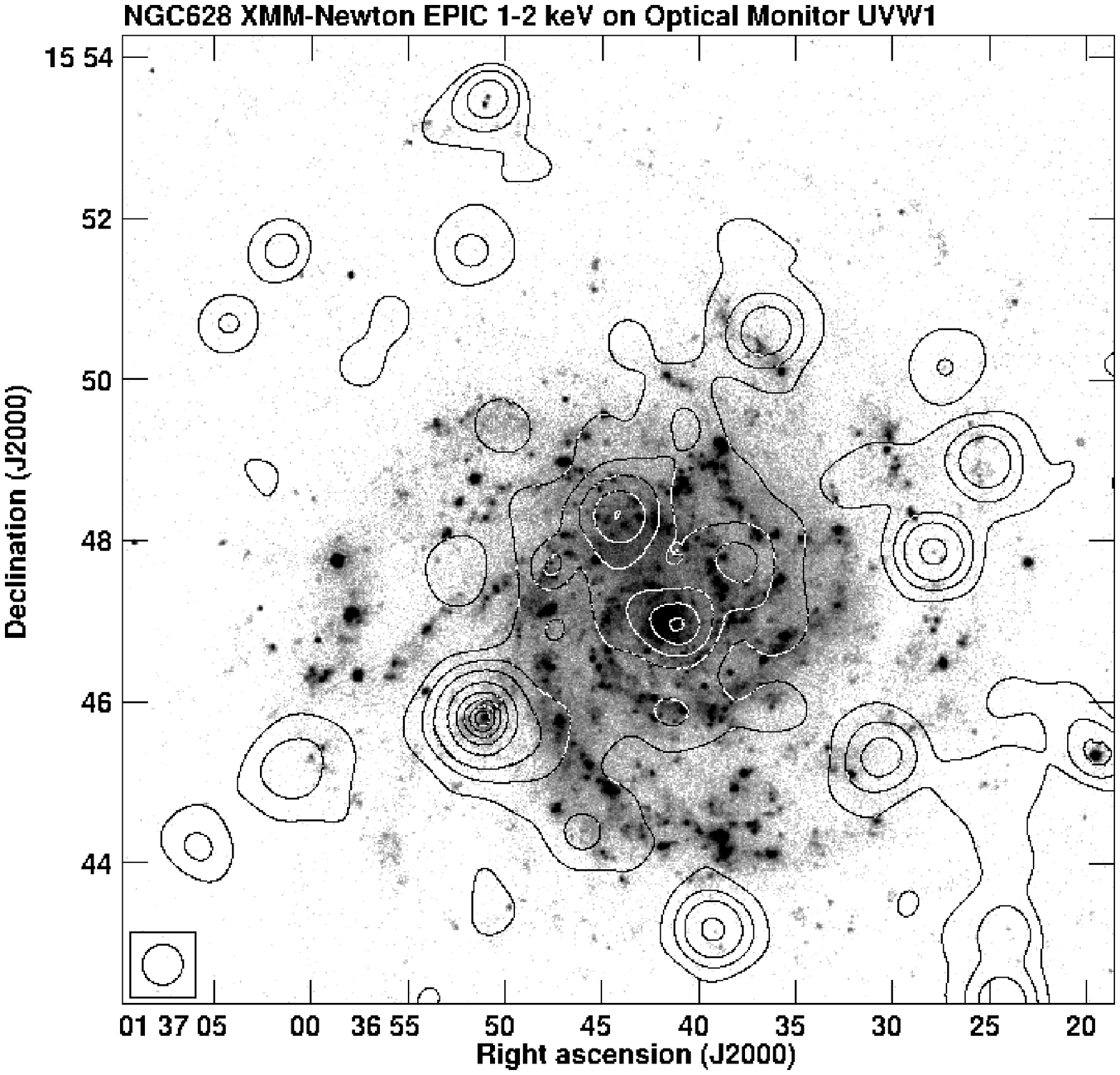}}
                \caption{
                {\it } Map of the soft X-ray emission from NGC\,628 in the 0.2 - 1 keV band overlaid on an H$\alpha$ image (left). The
                contours are 3, 5, 8, 16, 32, 64, 128, and 256 $\times$ rms. The map is adaptively smoothed with the
                largest scale of 30$\arcsec$. {\it Right:} Map of the medium X-ray emission from NGC\,628 in the 1 - 2 keV band overlaid on a UVW1 image.
                The contours are 3, 5, 8, 16, 32, 64, 128, and 256 $\times$ rms. The map is adaptively smoothed with the largest scale of 30$\arcsec$.
                }
                \label{xmaps}
        \end{figure*}

\begin{figure}[ht]
                        \resizebox{\hsize}{!}{\includegraphics[clip]{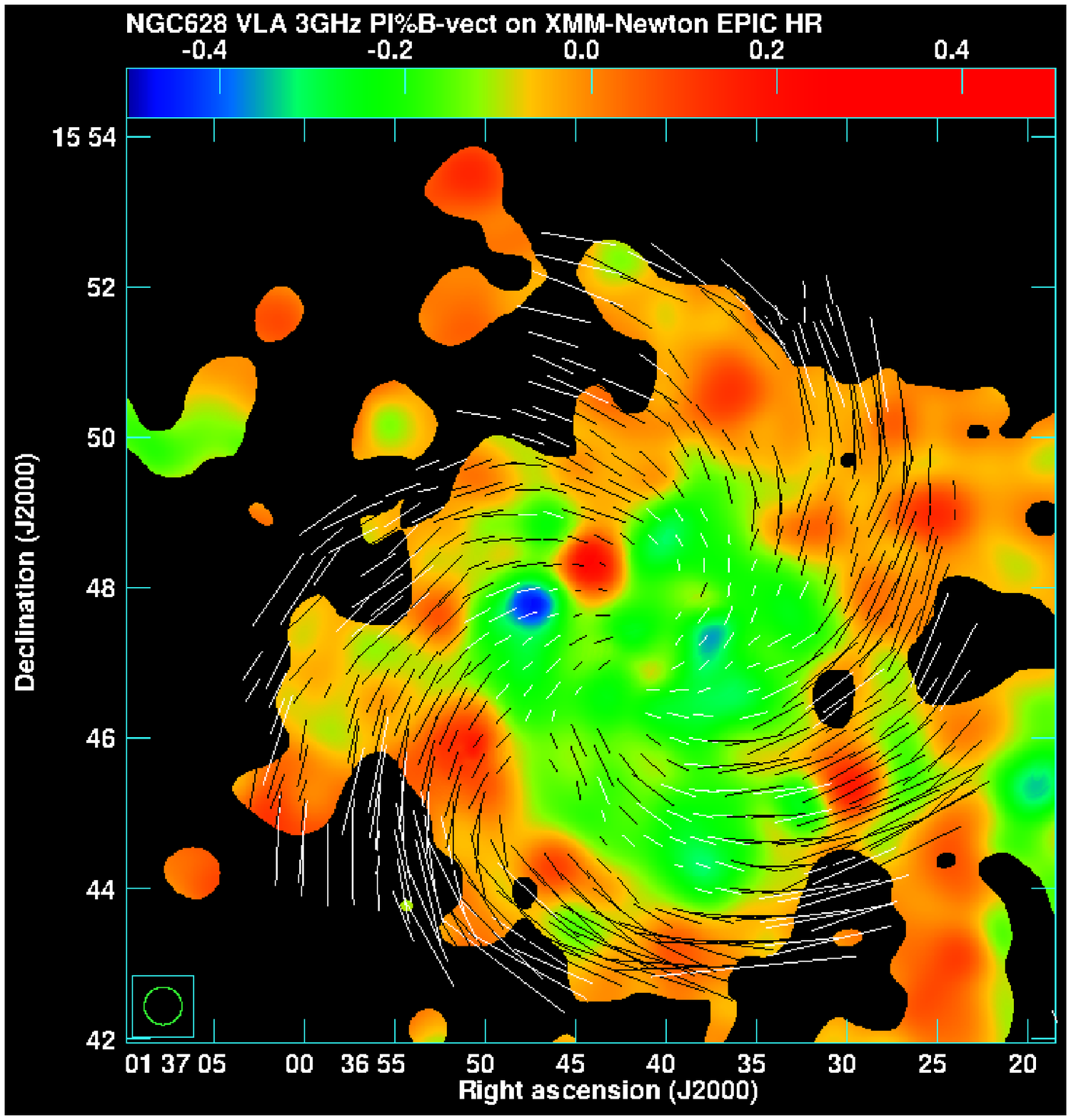}}
                \caption{Map of the hardness ratio of the medium and soft X-ray emission from NGC\,628 (Fig.~\ref{xmaps}), overlaid with 
                lines of the magnetic field proportional to the degree of polarisation. 
                The map is truncated at the 3$\sigma$ level of the soft X-ray emission map. 
                }
                \label{m74hr}
        \end{figure}

The X-ray emission from the hot gas is relatively faint and is mainly limited to the star-forming disk of NGC\,628 (Fig.~\ref{xmaps}). 
The diffuse emission is visible at a level not greater than 8$\sigma$ in the central part of the disk in the soft-band (0.2-1\,keV) image 
(Fig.~\ref{xmaps}, left) or 3-5$\sigma$ in the medium-band (1-2\,keV) image (Fig.~\ref{xmaps}). The hardness ratio (HR) map (Fig.~\ref{m74hr}) 
shows that the softest emission comes from the central part of the disk, which matches the brightest emission visible in the soft-band image 
(Fig.~\ref{xmaps}, left) very well. The harder emission in the outer parts of the disk can be associated with the hotter gas, possibly in the area of the magnetic arms.
The overall low level of the X-ray emission, with its softest part mainly coming from the central disk, agrees with the relatively low level of 
star formation in NGC\,628 and with the scenario that the star-forming activity has decreased over the last several hundred million years, 
which is a cooling time of the hot medium at temperatures of 0.3-0.6\,keV \citep[e.g.][]{wezgowiec20}.

\subsection{Spectral analysis of the X-ray emission}
\label{spectra}

\begin{figure*}[ht]
\resizebox{0.5\hsize}{!}{\includegraphics[clip]{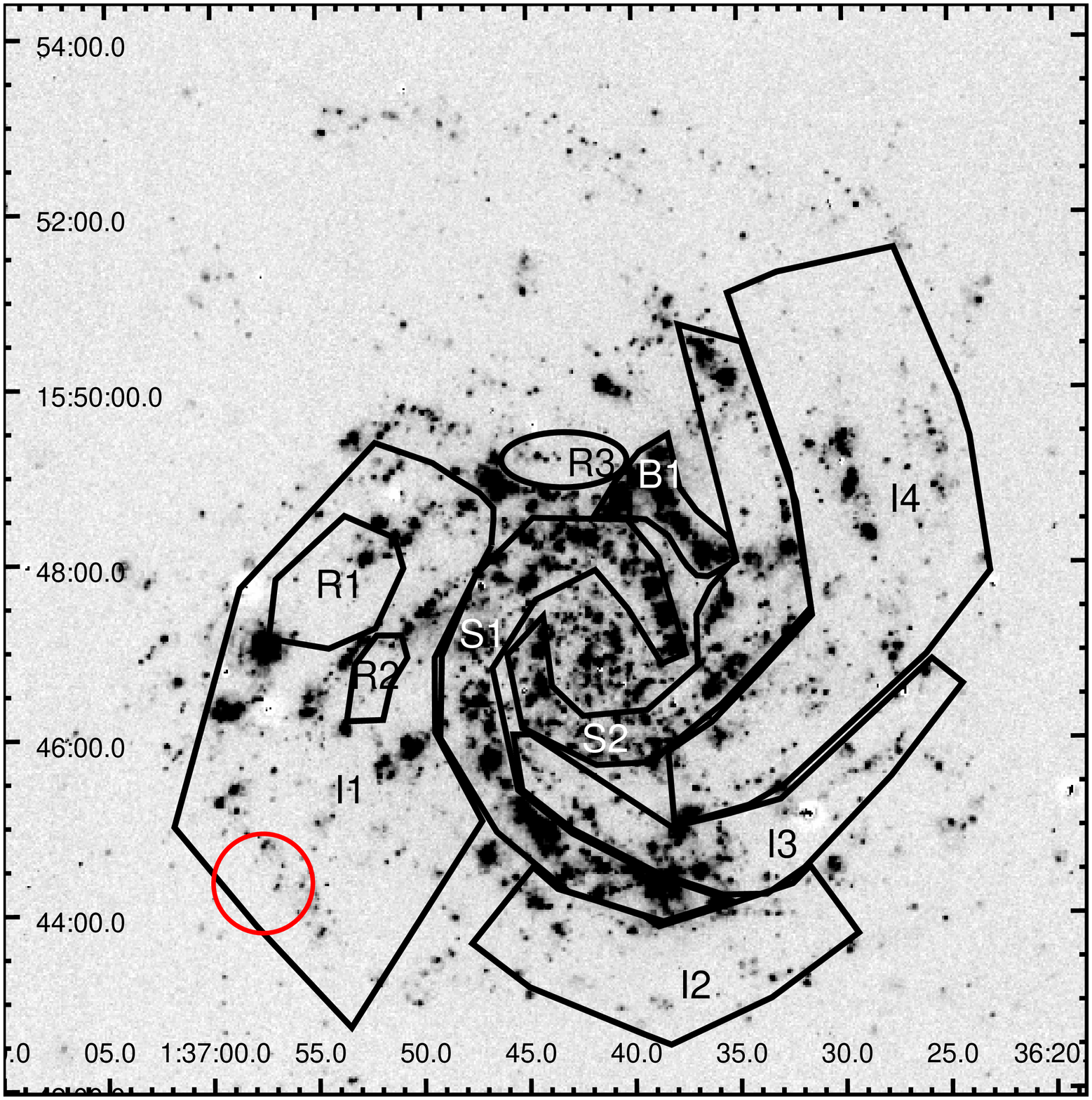}}
\resizebox{0.5\hsize}{!}{\includegraphics[clip]{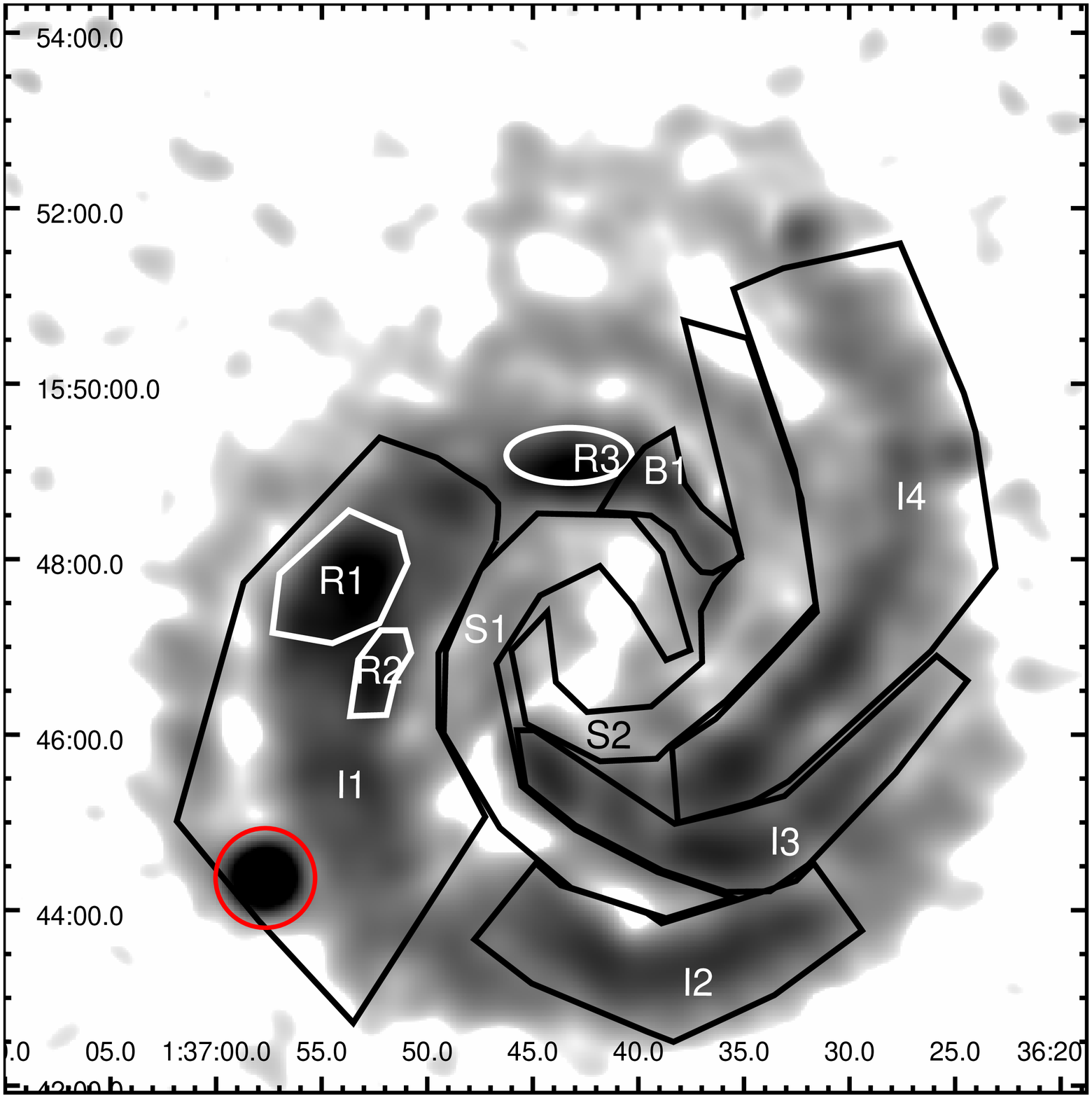}}

\caption{
        Regions of diffuse X-ray emission from NGC\,628 (see text for a detailed description) overlaid on the H$\alpha$ map (left) and on the polarised
        radio intensity at 3\,GHz (right). The region of the polarised radio source that we excluded from the analysis is marked with the red circle.
        }
\label{xregs}
\end{figure*}

Similar to our previous study \citep{wezgowiec16,wezgowiec20}, we investigated the properties of the hot gas in all regions 
that are prominent in the radio maps, that is, magnetic arms (highly polarised inter-arm regions) and spiral arms, as well as the halo above them.
The region selection was performed with the use of the polarised intensity map (right panel of Fig.~\ref{10radio}) and the H$\alpha$ map, 
tracing the ordered magnetic fields and star-forming regions, respectively. All regions are presented in Fig.~\ref{xregs} and Table~\ref{names}. 
In addition to the areas corresponding to the spiral arms (S) and the inter-arm spaces (I), we also selected other interesting regions of the galaxy 
that are associated with the largest \ion{H}{II} complexes (B1) or the bright polarised radio ridges (R1-R3). As mentioned in Sect.~\ref{xrayred}, 
all detected X-ray point sources were excluded from all regions. Additionally, an area of the strong polarised radio source 
(NVSS\,J013657+154422) was excluded from region I1 (red circle in Fig.~\ref{xregs}).

\begin{table}[ht]
\caption{\label{names}Regions of NGC\,628 that were used for the spectral analysis.}
\centering
\begin{tabular}{cl}
\hline\hline
Region name     & Region description      \\
\hline
\vspace{5pt}
B1               & large \ion{H}{II} region north of centre \\
\vspace{5pt}
I1               & eastern magnetic arm   \\
\vspace{5pt}
I2               & southern magnetic arm  \\
\vspace{5pt}
I3               & south-central magnetic arm  \\
\vspace{5pt}
I4               & western magnetic arm  \\
\vspace{5pt}
R1               & polarised feature within I1 (north)\\
\vspace{5pt}
R2               & polarised feature within I1 (south)\\
\vspace{5pt}
R3               & polarised feature north of centre \\
\vspace{5pt}
S1               & eastern spiral arm \\
\vspace{5pt}
S2               & western spiral arm \\
\hline
\end{tabular}
\end{table}

Following the extraction of spectra and the background subtraction described in Sect.~\ref{xrayred}, we performed spectral analysis for all regions to investigate the properties of the hot gas. Due to the slight contribution 
from the detector Al-K$\alpha$ line at 1.5\,keV, which is still visible in some of the spectra, 
the energy range 1.4-1.6\,keV was excluded prior to the model fitting for the best results. 
Despite the limited sensitivity of the X-ray data for NGC\,628, the spectra extracted from the regions of spiral arms (S1 and S2), as 
well as the large \ion{H}{ii} complex B1, were fitted with two thermal component models that account separately for the disk and the halo 
X-ray emission. The remaining inter-arm regions were fitted with only one thermal component to model the entire emission from the hot gas.
The thermal component used in our models is described by the {\it mekal} model \citep{mewe85,kaastra92} of thermal 
bremsstrahlung with additional emission lines. In all models, we also used a power-law component to account 
for the unresolved point sources and for the residual emission from the sources that were removed from the spectral regions 
(see Sect.~\ref{xrayred}). This additional component was not constrained in the case of the two-temperature model 
fitted to the spectrum from region B1, however. 
The absorption by the neutral hydrogen in our Galaxy is accounted for by the {\it wabs} model 
with a fixed value of the column density (Table~\ref{xdat}).

For the model fit to the S2 spiral arm region, 
the lower temperature of the two thermal components was fitted at the lower limit allowed by {\sc xspec12} (kT = 0.08\,keV).
Therefore, the real temperature of the hot gas can be lower. We also assumed symmetric uncertainties 
for this measurement because {\sc xspec12} does not allow for values lower than 0.08\,keV (see Table~\ref{xtabr}). 
The higher temperature in the same model fit was similar to that obtained for the inter-arm regions, for which a model with 
a single thermal component was fitted (Table~\ref{xtabr}). In our analysis of the very sensitive X-ray data for M\,83 \citep{wezgowiec20},
we found that the halo emission can be described with two thermal components
with temperatures of $\sim$0.1 and $\sim$0.3\,keV. Taking this into account, it is likely that both thermal components 
in the model fit to region S2 refer to the halo and that the emission from the hotter gas in the disk is not directly detected
in this moderately sensitive spectrum. Nevertheless, a relatively flat photon index of the power-law component in this model 
suggests that in fact only one thermal component is sufficient to model the spectrum of the emission from the spiral arm S2 (see Table~\ref{xtabr}). 
This is because in a low-sensitivity spectrum, the emission at higher energies is likely to be attributed either to the thermal 
or the power-law model component. This mixing of these two model components was described in \citet{wezgowiec16}.
The fact that for the spectrum from region B1 it was impossible to use the additional power-law component can also be due to this 
ambiguity, especially that the higher temperature is surprisingly high, even for hot disk gas, and at the same time, its uncertainty reaches 70\%.
The single thermal component model for the spectrum from B1 does allow for a constrained power-law component, but its 
uncertainty is relatively high, which only confirms the ambiguity mentioned before. Therefore, we argue that the contribution 
from the hot gas in the disk should be present in the spectrum from this bright star-forming region and only the sensitivity of the 
spectrum does not allow us to fit a model that includes both the second thermal and the power-law components simultaneously. Having 
this in mind, in the further analyses presented in this paper, we use the single thermal component model.

\begin{figure*}[ht]
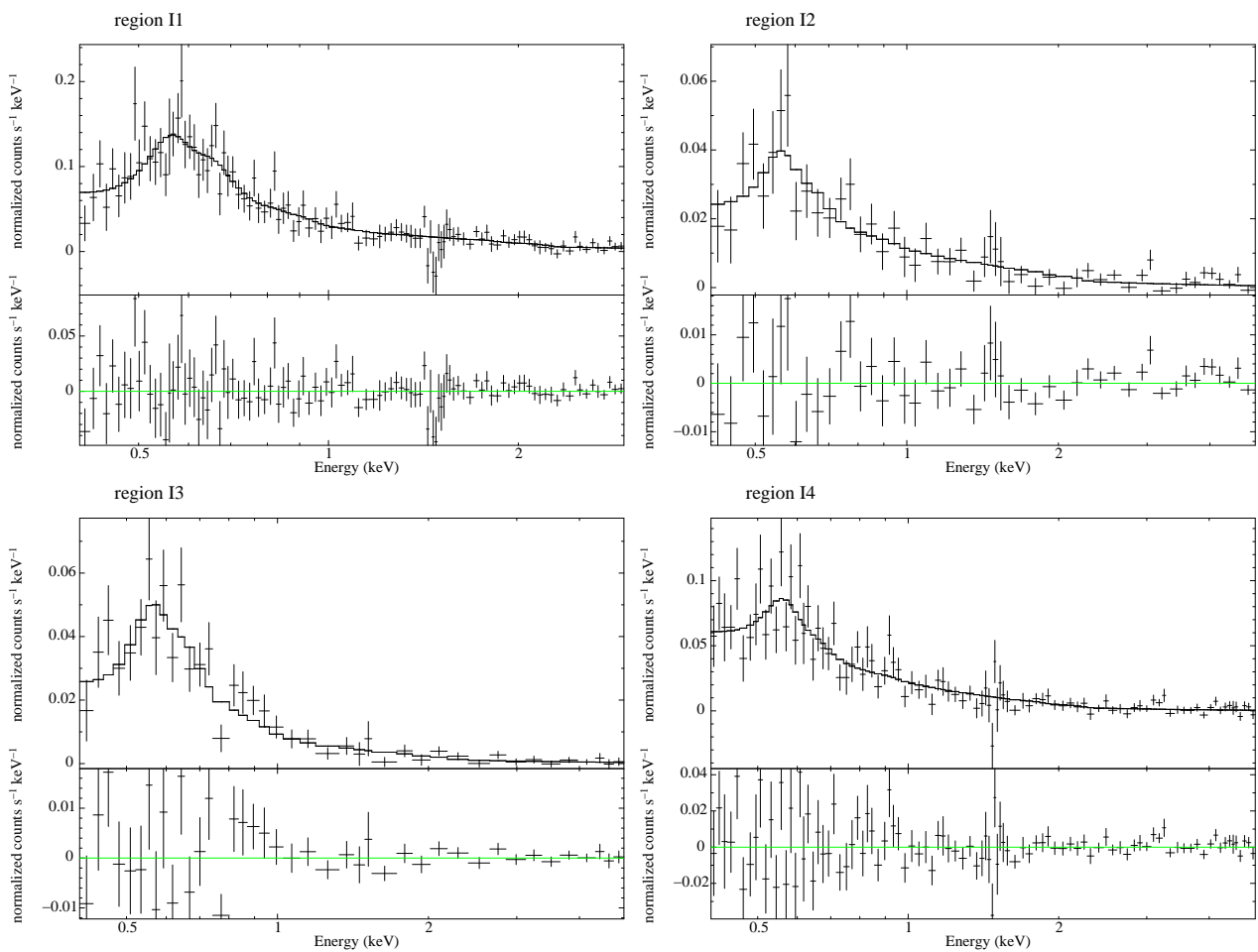

\begin{center}
\resizebox{0.45\hsize}{!}{\includegraphics[angle=-90]{m74_I1_normal_line.ps}}
\resizebox{0.45\hsize}{!}{\includegraphics[angle=-90]{m74_I2_normal_line.ps}}
\resizebox{0.45\hsize}{!}{\includegraphics[angle=-90]{m74_I3_normal.ps}}
\resizebox{0.45\hsize}{!}{\includegraphics[angle=-90]{m74_I4_normal_line.ps}}
\end{center}
\caption{Spectral model fits to the diffuse X-ray emission from the inter-arm regions of NGC\,628. See Tables~\ref{xtabr} and \ref{xfr}.}
\label{models}
\end{figure*}

\begin{figure}[ht]
\begin{center}
\resizebox{\hsize}{!}{\includegraphics[angle=-90]{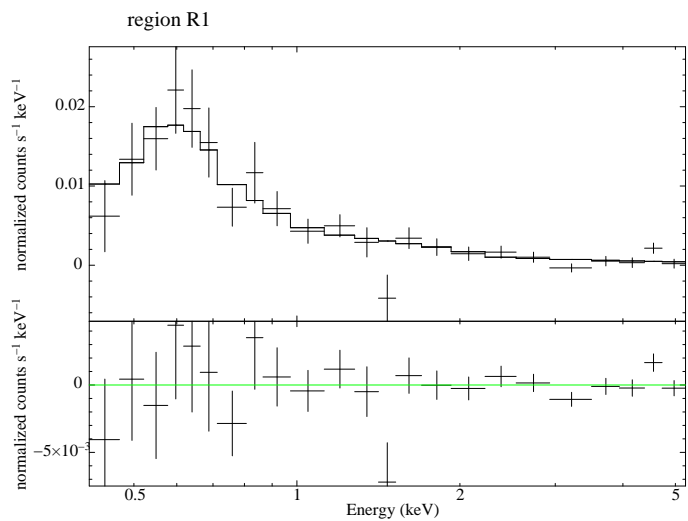}}
\end{center}
\caption{Spectral model fit to the diffuse X-ray emission from the polarised feature R1 within the inter-arm region I1. See Tables~\ref{xtabr} and \ref{xfr}.}
\label{R1}
\end{figure}

\begin{figure}[ht]
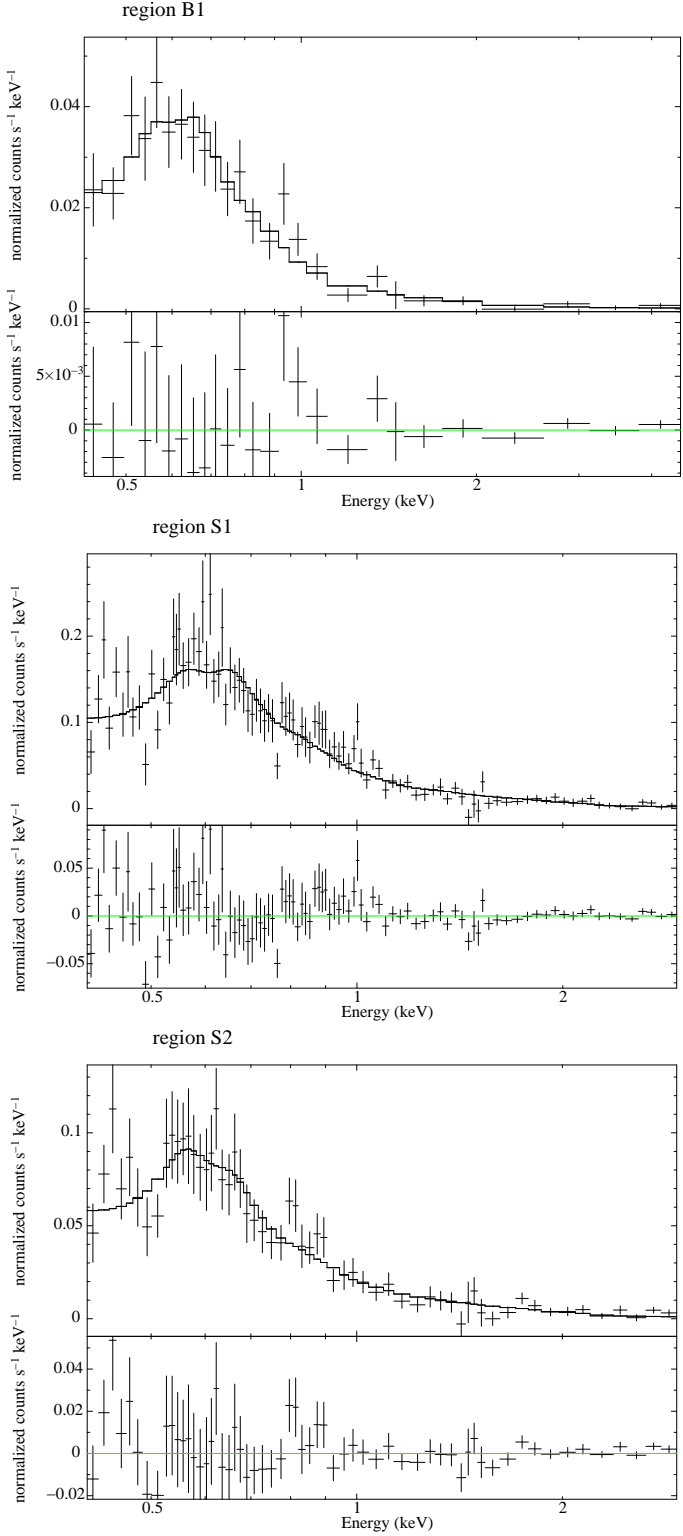

\begin{center}
\resizebox{\hsize}{!}{\includegraphics[angle=-90]{m74_B1_normal.ps}}
\resizebox{\hsize}{!}{\includegraphics[angle=-90]{m74_S1_normal_single.ps}}
\resizebox{\hsize}{!}{\includegraphics[angle=-90]{m74_S2_normal_single.ps}}
\end{center}
        \caption{Spectral model fits to the diffuse X-ray emission from the spiral arms of NGC\,628. See Tables~\ref{xtabr} and \ref{xfr}.}
\label{modelspirals}
\end{figure}

Only for the spectrum of the more pronounced spiral arm S1, which has significantly higher net counts and surface brightness
than that for the spiral arm S2 (see Table~\ref{xtabr}), was it possible to obtain a well-constrained model 
with two thermal components. Its temperatures of 0.20$\pm$0.02 and 0.77$\pm$0.17\,keV suggest a direct detection of the disk emission, 
being associated with the hotter component. For this spectrum, which is still moderately sensitive, 
it should still be possible to associate only one thermal component with the entire emission from the diffuse X-ray gas, whose parameters 
should be similar to those derived from the two thermal component model. We performed this test and found that in the single thermal 
component model, the global temperature is the average (weighted by normalisation) of the temperatures from the two thermal component model, and 
additionally, its power-law component is much better constrained than that in the two thermal component model. Consequently, 
the derived fluxes have lower uncertainties (see Table~\ref{xfr}), which results in better-constrained parameters of the hot gas.
Taking this into account, we decided to use the single thermal component model also for the spectrum from region S1. Therefore, 
in the following discussions of the properties of the hot gas for all regions, including the star-forming regions B1, S1, and S2, we use the single thermal component model fits to calculate the hot gas parameters. 
The parameters of the fitted models are presented in Table~\ref{xtabr}, and the derived fluxes are listed in Table~\ref{xfr}. 
The two thermal component models are marked in both tables with italics. 
The plots of the fitted (single thermal component) models with residuals are
presented in Figs.~\ref{models}--\ref{modelspirals}. The lowest sensitivity of the spectra was obtained for the regions 
of the polarised ridges R1, R2, and R3. As shown in Table~\ref{xtabr}, only for region R1 was it possible to obtain a constrained model fit.

\begin{table*}[ht]
        \caption{\label{xtabr} Model-fit parameters for the regions studied in NGC\,628.}
\centering
\begin{tabular}{clccccrr}
\hline\hline
Region& Model                               & kT$_1$                & kT$_2$        & Photon                         &$\chi_{\rm red}^2$ & Net       & Net cts\\
            & type                          & [keV]                 & [keV]         & Index                         &                   & cts       & /sqarcmin\\        
\hline
B1& wabs(mekal+power law)         & 0.23$^{+0.03}_{-0.02}$  & --                    & 2.22$^{+0.74}_{-1.12}$     &       0.91        & 1649      & 1994  \\
\vspace{5pt}
{\it B1}&{\it wabs(mekal+mekal)}  & {\it 0.22$^{+0.02}_{-0.03}$} & {\it 1.03$^{+0.75}_{-0.31}$}& --              &{\it 0.82}  & {\it 1649}& {\it 1994}\\
\vspace{5pt}
I1& wabs(mekal+power law)         & 0.20$\pm$0.01           & --                    & 1.39$^{+0.32}_{-0.31}$     &       1.01        & 6164      & 738   \\
\vspace{5pt}
I2& wabs(mekal+power law)         & 0.18$^{+0.04}_{-0.06}$  & --                    & 1.74$^{+0.56}_{-0.60}$     &       1.37        & 2134      & 738   \\
\vspace{5pt}
I3& wabs(mekal+power law)         & 0.20$\pm$0.02           & --                    & 1.76$^{+0.74}_{-0.72}$     &       1.15        & 2775      & 960   \\
\vspace{5pt}
I4& wabs(mekal+power law)         & 0.16$^{+0.03}_{-0.04}$  & --                    & 2.22$^{+0.41}_{-0.46}$     &       1.04        & 3775      & 705   \\
\vspace{5pt}
R1& wabs(mekal+power law)         & 0.21$^{+0.03}_{-0.04}$  & --                    & 1.27$^{+0.63}_{-0.64}$     &       1.06        & 982       & 798   \\
S1& wabs(mekal+power law)         & 0.23$\pm$0.01           & --                    & 2.18$^{+0.28}_{-0.33}$     &       1.04        & 7996      & 1985  \\
\vspace{5pt}
{\it S1}&{\it wabs(mekal+mekal+power law)}& {\it 0.20$\pm$0.02}& {\it 0.77$\pm$0.17}& {\it 1.63$^{+0.62}_{-1.04}$}&{\it 1.11}  & {\it 7996}& {\it 1985}\\
S2& wabs(mekal+power law)         & 0.21$\pm$0.02           & --                    & 2.28$^{+0.47}_{-0.66}$     &       1.07        & 5128      & 1260  \\
\vspace{5pt}
{\it S2}&{\it wabs(mekal+mekal+power law)}&{\it 0.08$\pm$0.03}&{\it 0.26$^{+0.05}_{-0.02}$}&{\it 1.32$^{+0.60}_{-0.55}$}&{\it 0.73}&{\it 5128}&{\it 1260}\\
\hline
\end{tabular}
\end{table*}

In their two-temperature model fit to the emission from the entire galaxy, \citet{owen09} found two thermal components characterised by 
temperatures of 0.2\,keV and 0.65\,keV for the halo and disk, respectively. 
The temperatures obtained by us in the two thermal component model fits to the spectra from 
the spiral arm S1 and the \ion{H}{ii} complex B1 correspond well with these values. 
As mentioned before, in the low-sensitivity spectrum of the emission from the spiral arm S2, we may not have detected 
the disk emission directly, and both thermal model components describe the halo emission. This seems justified because typically, much more emission comes 
from the halo gas than from the disk \citep[e.g.][]{tuellmann06,wezgowiec20}. 
For all inter-arm regions, the single-temperature 
model fits resulted in temperatures of about 0.2\,keV, matching the halo temperature obtained by \citet{owen09}. 
Last but not least, from their model fit to the entire emission from NGC\,628, 
\citet{owen09} calculated that the flux ratio of the cool and hot components is higher than 5.4.
This can certainly explain difficulties in the detection of the hot disk component in the lower-sensitivity spectra from smaller regions of the galaxy.

The highest temperature in the inter-arm regions, although still within uncertainties, 
was that derived for the polarised ridge R1, which is a part of region I1. 
This suggests that the temperature of the hot gas within I1 and outside of R1 could be lower. In this case, the highest temperature will be that 
for the inter-arm region I3, which corresponds to the bright magnetic arm (Figs.~\ref{10radio} and~\ref{xregs}). 

\begin{table}[ht]
        \caption{\label{xfr} Total (0.3 - 12 keV) unabsorbed fluxes in 10$^{-14}$erg\,cm$^{-2}$s$^{-1}$ for the modelled regions in NGC\,628.}
\centering
\begin{tabular}{ccccc}
\hline\hline
Region       & mekal 1                     & mekal 2                    & power law                 & total                      \\
\hline
B1           & 0.91$^{+0.30}_{-0.32}$      & --                         & 1.05$^{+3.48}_{-0.55}$    & 1.96$^{+3.78}_{-0.87}$     \\
\vspace{5pt}
{\it B1}     & {\it 1.10$^{+0.18}_{-0.26}$}& {\it 0.32$^{+0.45}_{-0.19}$} & --                      & {\it 1.42$^{+0.62}_{-0.46}$}\\
\vspace{5pt}
I1           & 3.70$^{+0.85}_{-0.78}$      & --                         & 12.80$^{+11.12}_{-5.67}$  & 16.5$^{+11.97}_{-6.46}$    \\
\vspace{5pt}
I2           & 0.65$^{+0.46}_{-0.50}$      & --                         & 3.12$^{+5.27}_{-1.63}$    & 3.76$^{+5.72}_{-2.13}$     \\
\vspace{5pt}
I3           & 1.50$\pm$0.50               & --                         & 2.60$^{+6.21}_{-1.48}$    & 4.10$^{+6.70}_{-1.97}$     \\
\vspace{5pt}
I4           & 1.43$^{+1.41}_{-0.99}$      & --                         & 5.36$^{+3.01}_{-1.39}$    & 6.79$^{+4.42}_{-2.38}$       \\
\vspace{5pt}
R1           & 0.49$^{+0.25}_{-0.26}$      & --                         & 2.46$^{+7.70}_{-1.78}$    & 2.95$^{+7.96}_{-2.04}$       \\
S1           & 3.53$^{+0.65}_{-0.63}$      & --                         & 6.53$^{+2.68}_{-1.43}$    & 10.07$^{+3.32}_{-2.06}$    \\
\vspace{5pt}
{\it S1}     & {\it 4.05$^{+1.00}_{-1.08}$}& {\it 1.14$^{+0.60}_{-0.51}$}& {\it 6.02$^{+33.89}_{-4.06}$}&{\it 11.20$^{+35.49}_{-5.65}$}\\
S2           & 1.68$^{+0.64}_{-0.59}$      & --                         & 3.03$^{+2.73}_{-0.83}$    & 4.71$^{+3.37}_{-2.25}$      \\
\vspace{5pt}
{\it S2}     & {\it 3.06$^{+7.14}_{-2.20}$}& {\it 2.18$^{+0.66}_{-0.69}$}& {\it 5.26$^{+12.80}_{-3.55}$}& {\it 10.5$^{+20.57}_{-6.45}$}\\
\hline
\end{tabular}
\end{table}

\section{Discussion}
\label{disc}

\subsection{Parameters of the hot gas}
\label{gasparams}

We used the model of thermal cooling and ionisation equilibrium of \citet{nulsen84} and the model-fit parameters 
to calculate the physical properties of the hot gas. In the ionisation equilibrium model, 
where $L_X=1.11\cdot \Lambda(T)\,n^2_e\,V\,\eta$, $\Lambda(T)$ is a cooling coefficient of about 
$10^{-22}\,{\rm erg}\,{\rm cm}^3\,{\rm s}^{-1}$ for temperatures of a few million K \citep{raymond76}, 
$n^2_e$ is number density, $V$ is the emitting volume, and $\eta$ is an unknown filling factor.
The emitting volume $V$ was calculated as a product of the surface of the spectral region and the height above 
the disk plane. We assumed a 10\,kpc halo extending above a 1\,kpc thick disk. The former was assumed as a typical extent of a gaseous 
halo in normal star-forming spiral galaxies \citep[e.g.][]{tuellmann06}. These volumes and the normalisations of our 
model fits together with the derived fluxes allowed us to obtain number densities, gas masses, and thermal energies of the hot gas.
The calculated parameters of the hot gas are presented in Table~\ref{global}. 

\begin{table}[ht]
        \caption{\label{global}Derived parameters of the hot gas from the studied regions in NGC\,628.}
\centering
\begin{tabular}{ccccc}
\hline\hline
Reg.& n\,$\eta^{-0.5}$       & M$_{gas}\,\eta^{0.5}$    & E$_{th}\,\eta^{0.5}$  & $\epsilon_{th}\,\eta^{-0.5}$ \\
\vspace{5pt}& [10$^{-3}$cm$^{-3}$]& [10$^6$M$_\odot$]   & [10$^{54}$\,erg]      & [10$^{-12}$\,erg\,cm$^{-3}$]                          \\
\hline
\vspace{5pt}
B1  & 1.55$^{+0.21}_{-0.26}$& 1.54$^{+0.21}_{-0.26}$& 1.01$^{+0.29}_{-0.25}$& 0.86$^{+0.25}_{-0.21}$  \\
\vspace{5pt}
I1  & 1.05$^{+0.08}_{-0.09}$ & 10.5$^{+0.08}_{-0.09}$   & 6.00$\pm$0.76                & 0.50$\pm0.06$                \\
\vspace{5pt}
I2  & 0.78$^{+0.16}_{-0.23}$ & 2.69$^{+0.54}_{-0.80}$   & 1.39$^{+0.65}_{-0.74}$& 0.34$^{+0.16}_{-0.18}$       \\
\vspace{5pt}
I3  & 1.14$^{+0.12}_{-0.17}$ & 3.94$^{+0.42}_{-0.58}$   & 2.26$^{+0.49}_{-0.52}$& 0.55$^{+0.12}_{-0.13}$       \\
\vspace{5pt}
I4  & 0.91$^{+0.25}_{-0.23}$ & 5.81$^{+1.62}_{-1.46}$   & 2.66$^{+1.38}_{-1.17}$& 0.35$^{+0.18}_{-0.15}$       \\
\vspace{5pt}
R1  & 0.97$^{+0.18}_{-0.24}$ & 1.42$^{+0.26}_{-0.35}$   & 0.86$^{+0.30}_{-0.33}$& 0.49$^{+0.17}_{-0.19}$       \\
\vspace{5pt}
S1  & 1.40$^{+0.11}_{-0.12}$& 6.74$^{+0.52}_{-0.58}$& 4.44$^{+0.55}_{-0.56}$& 0.77$\pm$+0.10\\
\vspace{5pt}
S2   & 1.23$^{+0.17}_{-0.20}$& 3.87$^{+0.54}_{-0.63}$& 2.33$^{+0.58}_{-0.57}$& 0.40$\pm$0.10\\
\hline
\end{tabular}
\tablefoot{
        The columns are the region name, electron number density, total gas mass, total thermal energy, thermal energy density, and cooling time.
        $\eta$ is the volume-filling factor.
        }
\end{table}

\subsection{Magnetic fields in NGC\,628}
\label{bfields}

The magnetic fields of NGC\,628 were thoroughly studied by \citet{mulcahy17}, therefore we only focus on the strengths 
of the total and regular component and on energy densities of the magnetic fields in our spectral regions. At the frequency of 3\,GHz, 
the thermal radio emission can still be significant and should be taken into account in calculations of the strengths of the magnetic fields. 
Following the calculations of \citet{mulcahy17}, for the star-forming regions, that is, B1, S1, and S2, we used a thermal fraction of 20\%. 
For the areas of the inter-arm regions and for the polarised ridges, where no significant star formation is visible, 
we simultanesouly used the entire radio emission as non-thermal. 

To calculate the strengths and energy densities of the magnetic field, we also followed 
\citet{mulcahy17} and used the revised formula for the energy equipartition provided by \citet{beck05}.  
In addition to the non-thermal intensities, we used a path length through the source of $L = 1\,{\rm kpc}$, which is the thickness of the synchrotron disk, 
and also took its inclination of 7$\degr$ into account. Following \citet{mulcahy17}, we adopted a synchrotron spectral index of $\alpha_{\rm n}=-1$ 
and a ratio of cosmic-ray proton to electron number densities of $K=100$. 
Following the argumentation of these authors, we note that because the strength of the total magnetic field scales with synchrotron 
intensity $I_{nth}$ as 

\begin{equation}
B_{tot} = (I_{nth}/(K + 1)L)^{1/(3-\alpha_{\rm n}}), 
\end{equation}

\noindent
realistic uncertainties in $L$ and $K$ of a factor of two lead to a magnetic field strength variation of about 20\%. A change in the non-thermal 
spectral index of about 30\% does not affect the result by more than 5\% either. 
The uncertainties of the non-thermal intensity mostly depend on the thermal fraction, which can change from 0 to 40\%
\citep[see Fig. 11 in][]{mulcahy17} within each star-forming region mentioned before. Because for these regions we assumed 
a thermal fraction of 20\%, we also changed the non-thermal intensities (for consistency also of the inter-arm regions) by 25\% to account for these 
local variations of the thermal fraction. 
Nevertheless, this change in intensity alters the strength of the magnetic field by less than 10\%.
Last but not least, the uncertainties in $K$ and $L$ can be considered systematic errors because they affect 
the calculations for all regions in the same manner. Because our aim is to compare the properties of the magnetic field (and of the hot gas) 
between different regions of M\,74, these uncertainties would not be relevant. Therefore, we took the uncertainties 
in non-thermal intensity and spectral index into account and assumed an error of 10\% for the strengths and 20\% for the energy densities 
of the magnetic field, because $\epsilon_B\,\propto\,B^2$.

\begin{table}[ht]
        \caption{\label{magparams} Properties of the magnetic fields in NGC\,628 derived from the maps presented in Fig.~\ref{10radio}.}
\centering
\begin{tabular}{cccccc}
\hline\hline
Reg.            &S$_{synch}$    &p$_{synch}$    &B$_{tot}$     &$\epsilon_B$     &B$_{ord}$   \\
                &[mJy           &[\%]           &[$\mu$G]      &[10$^{-12}$      &[$\mu$G]     \\
                &/beam]         &               &              &erg\,cm$^{-3}$]  &              \\
\hline
\vspace{5pt}
B1              &  1.17         &    7.5        &  12.8$\pm$1.3&    6.5$\pm$1.3  &  3.4$\pm$0.3  \\
\vspace{5pt}
I1              &  0.46         &   19.9        &  9.9$\pm$1   &    3.9$\pm$0.8  &  4.4$\pm$0.4  \\
\vspace{5pt}
I2              &  0.21         &   40.8        &  7.7$\pm$0.8 &    2.4$\pm$0.5  &  5.1$\pm$0.5  \\
\vspace{5pt}
I3              &  0.52         &   17.3        &  10.2$\pm$1  &    4.2$\pm$0.8  &  4.2$\pm$0.4  \\
\vspace{5pt}
I4              &  0.30         &   22.4        &  8.8$\pm$0.9 &    3.1$\pm$0.6  &  4.2$\pm$0.4  \\
\vspace{5pt}
R1              &  0.70         &   19.8        &  10.9$\pm$1.1&    4.8$\pm$1    &  4.8$\pm$0.5        \\
\vspace{5pt}
R2              &  0.58         &   23.9        &  10.3$\pm$1  &    4.3$\pm$0.9  &  5.1$\pm$0.5  \\
\vspace{5pt}
R3              &  0.82         &   16.9        &  11.5$\pm$1.2&    5.2$\pm$1    &  4.6$\pm$0.5        \\
\vspace{5pt}
S1              &  0.71         &    6.2        &  11.3$\pm$1.1&    5.1$\pm$1    &  2.7$\pm$0.3        \\
\vspace{5pt}
S2              &  0.41         &    9.4        &   9.8$\pm$1  &    3.8$\pm$0.8  &  2.9$\pm$0.3  \\
\hline
\end{tabular}
\tablefoot{
The columns are the region name, the non-thermal radio flux, the degree of polarisation, the total magnetic field strength, the energy density of the magnetic field, 
        and the ordered magnetic field strength.
}
\end{table}

The calculated strengths of the magnetic fields for all studied regions are in the range of 8-13\,$\mu$G, which agrees with the values 
obtained by \citet{mulcahy17}. The lowest strengths of the magnetic field of 8.2-10\,$\mu$G were obtained for the inter-arm regions 
hosting the magnetic arms, where little or no star formation is observed. The highest strengths of the magnetic field were derived 
for the largest \ion{H}{ii} complexes in region B1, and quite surprisingly, for the area of the polarised ridge R3. The highest strengths 
of the ordered magnetic fields are found in the magnetic arm regions and in the embedded polarised ridges, reaching 
5.1$\pm$0.3\,$\mu$G in region R2. Nevertheless, an equally strong ordered magnetic field can be found in the inter-arm region I2, which is located 
in the outer part of the southern galactic disk.

\subsection{Hot gas and magnetic fields}
\label{hotfields}

The analysis of the X-ray data for NGC\,628 provided the global parameters of the entire hot gas in this galaxy, 
that is, the disk and the halo. 
Correspondingly, the observed radio emission was integrated along the line of sight and therefore also brought 
 information about the whole galaxy. 

Prior to a direct comparison of the densities of energy stored in the hot gas and the magnetic fields, we needed to take into account that 
the derived parameters of the hot gas are related to an unknown filling factor ($\eta$) that likely varies between the analysed regions, 
especially the spiral arms and the inter-arm areas. 
Because it is difficult to provide reliable values of the filling factors, we introduced a new quantity, energy per particle $E_p$, which is the ratio of the
thermal energy density and the electron density of the hot gas. In this ratio, the dependence on the filling factor is removed, which allows measuring the energy stored in the hot gas directly \citep{wezgowiec16,wezgowiec20}. 

Table~\ref{particles} and Fig.~\ref{epmag} show the relation between the (hot gas) energy per particle and the energy density 
of the magnetic fields. While all regions are labelled in the plots, 
the regions of the spiral arms and the inter-arm regions are marked with black and red, respectively.
Because of the large uncertainties of energies per particle, caused by the low sensitivity of the X-ray 
spectra and the resulting large uncertainties of the fitted model normalisations, it is difficult to determine any possible differences between 
the regions of the spiral and the magnetic arms of NGC\,628. Nevertheless, it seems that in contrast to our previous findings for 
NGC\,6946 \citep{wezgowiec16} and M\,83 \citep{wezgowiec20}, a decrease in the energy density of the magnetic fields 
in the magnetic arms is not accompanied by an increase in the energy of the hot gas. 
The ordering of the magnetic field (Fig.~\ref{epord}), however, is higher in the magnetic arms than in the spiral arms, 
as observed before in NGC\,6946 \citep{wezgowiec16} and M\,83 \citep{wezgowiec20}. It resembles the former galaxy more because 
similarly to NGC\,6946, the distribution of the star-forming regions in NGC 628 presents a high contrast between the spiral arms and the inter-arm regions.
As we suggested in \citet{wezgowiec16}, this contrast may have an important influence on the formation of the prominent magnetic arms.

An interesting phenomenon is observed when the relation between energies per particle and energy densities of the magnetic field is compared 
to that obtained for M\,83 \citep{wezgowiec20}. Surprisingly, while both galaxies show very similar energies per particle, 
the derived energy densities of the magnetic fields in the regions of M\,83 are twice higher. 
The difference between the energy densities of the magnetic fields can easily be explained with the star-forming 
activities of the two galaxies. M\,83 is a starburst galaxy with a very high gas mass density \citep{lundgren04}, while the star-forming 
activity of NGC\,628 currently is at a moderate to low level \citep{marcum01,kennicutt11}. 
This is further supported by the overall low level of the X-ray emission in the presented maps of NGC\,628 (see Sect.~\ref{dist}).
If the energies stored in the hot gas were depending only on star formation, the energies per particle should also be higher in M\,83 than in NGC\,628. 
Because they are comparable, a contribution from a source of heating that is not directly related to star formation is needed to explain 
the increase in the energy of the hot gas in NGC\,628. We discuss this in more detail in Sect.~\ref{heat}.

\subsection{Signs of interaction?}
\label{interact}

\begin{figure}[ht]
\begin{center}
\resizebox{\hsize}{!}{\includegraphics[]{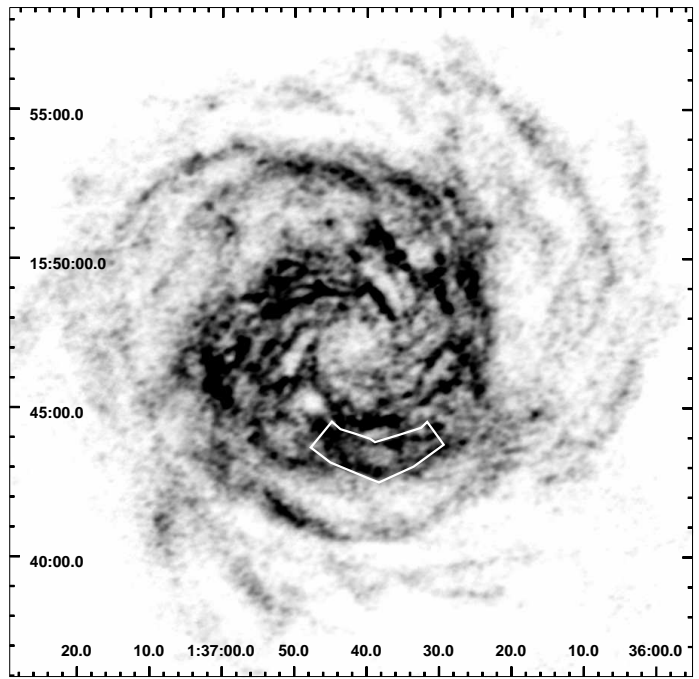}}
\end{center}
        \caption{\ion{H}{i} map of NGC\,628 from The \ion{H}{i} Nearby Galaxy Survey \citep[THINGS;][]{walter08}. Region I2 is marked in white.}
\label{ngc628hi}
\end{figure}

\begin{figure}[ht]
\resizebox{\hsize}{!}{\includegraphics[clip,angle=-90]{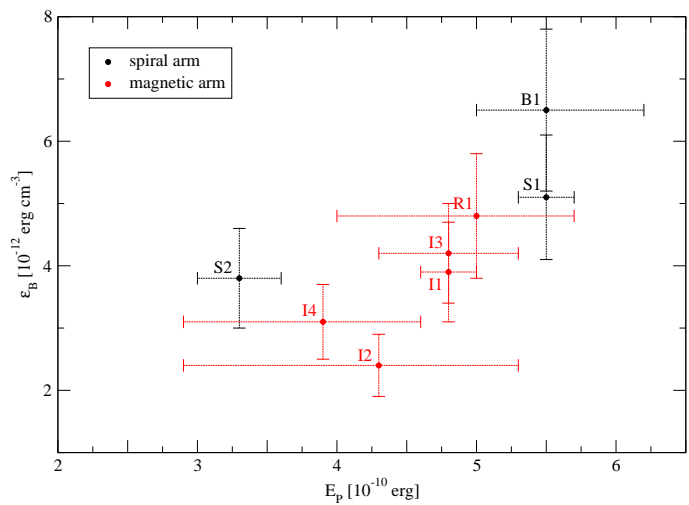}}
        \caption{Relation of the energy per particle (${\rm E}_{\rm P}$) and the energy density of the magnetic field (${\rm\epsilon}_{\rm B}$) for 
        regions associated with the spiral (black) and the magnetic (red) arms of NGC\,628. 
        }
\label{epmag}
\end{figure}

\begin{figure}[ht]
\resizebox{\hsize}{!}{\includegraphics[clip,angle=-90]{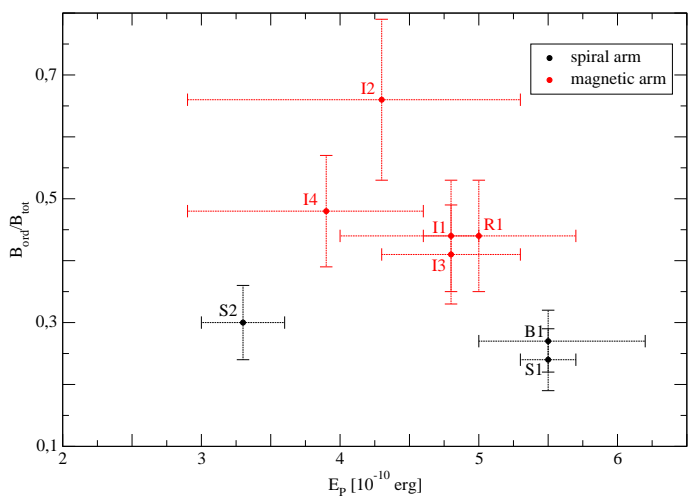}}
        \caption{Relation of the energy per particle (${\rm E}_{\rm P}$) and order (${\rm B}_{\rm ord}/{\rm B}_{\rm tot}$) 
        of the magnetic field for regions associated with the spiral (black) and the magnetic (red) arms of NGC\,628.
        }
\label{epord}
\end{figure}

We studied the inter-arm region I2 in the outer southern disk of NGC\,628 in more detail. The isolation of this galaxy and 
that it has not interacted with the close group companions for at least 1\,Gyr (see Sect.~\ref{recon}) appears to be doubtful. The magnetic arm in region I2 does not 
smoothly follow the spiral pattern of the galaxy, but is significantly bent northwards (right panel of Fig.~\ref{xregs}). 
This region shows the lowest energy density of the magnetic field ($\epsilon_B$) 
and the highest ordering of the magnetic field (exceeding 60\%, Fig.~\ref{epord}) of all regions 
we studied in our analyses. 
Furthermore, as mentioned in Sect.~\ref{radio}, the sensitive D-configuration map allowed us to trace the prominent western magnetic arm 
already to the north of the galaxy. At the same time, while the spiral and magnetic arms are usually observed alternately, 
the separated magnetic arms I3 and I4 (right panel of Fig.~\ref{xregs}) look just like one magnetic arm that was split in the past. 
This was also the conclusion of the analysis performed by \citet{mulcahy17}. Last but not least, the southern spiral arm (S1) looks 
well defined and follows the spiral structure of the galaxy until the southern position at which the magnetic arm I2 and I3 meet. 
S1 broadens slightly in this area and seems truncated (Fig.~\ref{xregs}). All these features suggest the possibility of past interaction 
of NGC\,628 with a companion galaxy, during which the currently southern spiral arm was disturbed and the magnetic arms in this part 
of the galaxy were pulled-out and stretched. The resulting shear of the magnetic field caused the observed high order and low energy density 
of the magnetic field in region I2. The latter, together with the lack of an increase in the temperature of the hot gas in this region, rather excludes 
the possibility of compression induced by ram-pressure. This tidal interaction scenario would resemble 
the very similar case of the Virgo cluster spiral galaxy NGC\,4254 \citep{wezgowiec12}. 
The shear of the magnetic field is further supported by the fact that region I2 is located directly at the 
sharp gradient of the \ion{H}{i} emission (Fig.~\ref{ngc628hi}).

\subsection{Additional heating of the hot gas}
\label{heat}

While it is widely accepted that the hot X-ray gas in spiral galaxies is produced in star formation and heated via supernova shocks, other mechanisms might also allow the ISM to reach X-ray temperatures, one of which is cosmic-ray streaming \citep{heintz20}. 
This mechanism, however, is expected to heat gas at densities of about 10$^{-3}$cm$^{-3}$ (derived from our spectral fits) to temperatures 
of several 10$^5$\,K or about 0.05\,keV. This is significantly lower than the temperatures of the hot gas observed in NGC\,628 (Table~\ref{xtabr}).

As we mentioned in Sect.~\ref{spectra}, our analysis of the X-ray data, as well as the results presented by \citet{owen09}, show that the majority 
of the diffuse X-ray emission comes from the hot gas in the halo of NGC\,628. \citet{mulcahy17} found signs of an efficient transport of the magnetic 
fields into the halo, that is, Parker instabilities and super-bubbles. Their Faraday rotation and depolarisation study also suggest the presence 
of vertical magnetic fields. The most prominent area was found to be our magnetic arm I1. If these fields form reversing loops like that 
found in NGC\,4631 \citep{mora19}, magnetic reconnection is possible. One of its effects is the heating of the surrounding gas.

In the reconnection heating scenario discussed by us for NGC\,6946 \citep{wezgowiec16} and M\,83 \citep{wezgowiec20}, magnetic reconnection effects 
convert energy of the magnetic fields into energy of the ISM gas. We note here that this conversion is a complex process that 
includes ohmic heating or the creation of high-frequency Alfv\'en waves, which in turn can cause anisotropic heating of ions \citep{lazarian99,vishniac99}. 
Furthermore, because the nature of turbulence in the reconnection zone is not clear, the rate of the energy emission is difficult to estimate. 
Therefore, in the following discussion we refer to 'reconnection heating' as a variety of effects that convert magnetic into thermal energy. 
Because they act more efficiently on the turbulent component 
(field reversals required), these effects also lead to a higher order of the magnetic field. 
The large-scale vertical magnetic fields in NGC\,628 and internal energies of the hot gas, being similar to that of a massive starburst M\,83, 
allow the possibility that the additional source of heating mentioned in Sect.~\ref{hotfields} 
is reconnection heating that occurs in the halo of NGC\,628.  
Especially interesting in this aspect is region R3, where a strong and ordered magnetic field was found by \citet{mulcahy17}. 
The authors argued that this polarisation peak could be due to compression of the magnetic field by an asymmetrically expanding \ion{H}{i} bubble 
caused by several supernovae. They also found a low Faraday depth at this position, which signifies a rather weak magnetic field along the line of sight, 
that is, the vertical component. A strong and ordered magnetic field in the sky plane over the region of around 1\,kpc, which is a typical wavelength 
of the Parker instability for equipartition conditions \citep{hanasz98}, and a corresponding weak 
line-of-sight component of the magnetic field can be well explained by magnetic reconnection that removes the vertical component of the magnetic 
field dragged by the expanding supernova bubbles. This would be done by reconnecting of the anti-parallel field lines around the adjacent bubbles 
and leaving the compressed magnetic field at their top. Unfortunately, the sensitivity 
of the X-ray data does not allow us to obtain a reliable value of the hot gas temperature in this area. A detection of an increased temperature 
in this region would provide further arguments for reconnection heating. 

The co-existence of hotter gas and the ordered magnetic fields seems to be crucial for the detection of magnetic reconnection effects. 
Because the shock-heated ISM can move to significant distances (up to 10\,kpc) in all directions \citep{hu19}, it can be present both 
in the inter-arm regions and in the galactic halo. Nevertheless, it is unlikely that this ISM will have higher temperatures and internal energies 
(our energy per particle, independent of the filling factor) than that found directly in the star-forming regions. Therefore, we propose here 
that for a reliable detection of magnetic reconnection effects, a 
higher temperature of the hot gas (and the corresponding energy per particle) and a simultaneous lower energy density of the magnetic field accompanied 
with its high order is crucial. Our studies of the radio and X-ray data for NGC\,6946 \citep{wezgowiec16}, M\,83 \citep{wezgowiec20}, and NGC\,628 presented 
in this paper suggest that only when all these three conditions are fulfilled can magnetic reconnection heating be distinguished 
from supernova shock heating. 

In this sense, in all inter-arm regions and polarised ridges (I1-4 and R1-R3), lower energy densities and higher order 
of the magnetic fields are observed than in the star-forming regions. However, the temperatures and energies per particle are 
rather constant throughout the galaxy when the significant uncertainties are taken into account. Nevertheless, we note here that 
this does not contradict the reconnection heating if most of its effects occur in the halo of NGC\,628, as proposed before. If present above the 
entire disk, it could cause similar temperatures of the hot gas throughout the galaxy and contribute to total energies per particle, which, as 
mentioned above, are the same as those for the starbursting galaxy M\,83. In this case, a similar contribution from the 
reconnection heating is also expected from the halo of M\,83. In particular, large-scale ordered magnetic fields in the 
disk-halo interface of M\,83 were suggested by the polarised radio observations by \citet{heald16}. Significant differences, however, are visible 
when we compare the densities of the hot gas in the haloes of the two galaxies. They are two to three times lower in the case of NGC\,628.
Although all number densities derived from the spectral model fits depend on an unknown filling factor, 
for the halo gas, it is justified to directly compare 
the obtained values,because at heights of 2.5\,kpc above the disk, the filling factor for the hot gas reaches unity \citep{deavillez00}.
Because magnetic reconnection is more efficient at lower number densities \citep{hanasz98}, this would suggest a higher contribution 
from reconnection heating in the halo of NGC\,628 and explain that the energies per particle are comparable to the energy found for M\,83. 

To estimate the reduction of the energy density of the magnetic field of NGC\,628 due to magnetic reconnection effects, we note here 
that in regions of star formation (B1, S1, and S2) the strength of the ordered magnetic field is lower than that of the total magnetic field 
by a factor of about three and in the remaining regions (inter-arm and polarised ridge) by a factor of about two (Table~\ref{magparams}).
This means that the energy density of the turbulent magnetic field is about 75-90\% of the total energy density. With total energy densities 
of 3-6$\cdot$10$^{-12}$\,erg\,cm$^{-3}$ (Table~\ref{magparams}), this means that energy densities of 2-5$\cdot$10$^{-12}$\,erg\,cm$^{-3}$ are stored 
in the turbulent component of the magnetic field. 
Because as much as 90\% of this energy can be lost during the expansion of the magnetic field into the halo \citep{hanasz98}, this leaves 
2-5$\cdot$10$^{-13}$\,erg\,cm$^{-3}$ for the conversion into thermal energy via the reconnection heating effects. If we assume their 
efficiency of 50\% \citep{hanasz98}, gas at densities of about 10$^{-3}$\,cm$^{-3}$ (in the inter-arm or polarised ridge regions, see 
Table~\ref{magparams}) will gain an additional energy of 1-2.5$\sim$10$^{-10}$\,erg per particle. This amounts to 20-50\% of the 
current energy per particle for the inter-arm regions (Table~\ref{particles}). Consequently, two or three times higher densities of the hot gas, like 
that found in M\,83 \citep{wezgowiec20}, would require two to three times higher energy densities of the magnetic field 
 to result in a similar increase in the energy per particle. As we mentioned before, the energy densities of the magnetic field in M\,83 
are roughly twice higher than that in M\,74. A higher efficiency of the reconnection heating therefore seems to be possible in the latter galaxy, 
especially in its regions of polarised radio ridges.

\begin{table}[ht]
\caption{\label{particles}Thermal energy per particle and energy densities of the magnetic field for the studied regions in NGC\,628.}
\centering
\begin{tabular}{ccc}
\hline\hline
Region      & $E_p$                 & $\epsilon_{B}$\\
\vspace{5pt}
            & [10$^{-10}$\,erg]     & [10$^{-12}$\,erg\,cm$^{-3}$]\\
\hline
\vspace{5pt}
B1          & 5.5$^{+0.7}_{-0.5}$   & 6.5$\pm$1.3  \\
\vspace{5pt}
I1          & 4.8$\pm$0.2           & 3.9$\pm$0.8  \\
\vspace{5pt}
I2          & 4.3$^{+1.0}_{-1.4}$   & 2.4$\pm$0.5  \\
\vspace{5pt}
I3          & 4.8$\pm$0.5           & 4.2$\pm$0.8  \\
\vspace{5pt}
I4          & 3.9$^{+0.7}_{-1.0}$   & 3.1$\pm$0.6  \\
\vspace{5pt}
R1          & 5.0$^{+0.7}_{-1.0}$   & 4.8$\pm$1    \\
\vspace{5pt}
S1          & 5.5$\pm$0.2           & 5.1$\pm$1    \\
\vspace{5pt}
S2          & 3.3$\pm$0.3           & 3.8$\pm$0.8  \\
\hline
\end{tabular}
\end{table}

\section{Summary and conclusions}
\label{cons}

We presented the results of our analyses of properties of the hot gas and magnetic fields in NGC\,628. 
The studies of the regions of spiral arm, inter-arm, polarised radio ridges, and the halo above them allowed us to 
draw the following conclusions:

\begin{itemize}
\item Overall, lower energy densities and higher order of the magnetic fields are found in the inter-arm regions and in areas 
        of the radio-polarised ridges compared to the star-forming spiral arms.
\item The temperatures of the hot gas and the resulting energies per particle are similar for all regions within the derived uncertainties.
\item The comparison of the above findings with that for the starburst galaxy M\,83 \citep{wezgowiec20} suggests that additional heating 
        of the hot gas in NGC\,628 may be present, which does not result directly from star formation. We propose that magnetic reconnection heating 
        occurs in the halo of NGC\,628 to explain this.
\item A more direct evidence of this halo reconnection might be the polarised radio ridges that show strong ordered magnetic fields and 
        were reported to be located in or above regions of weak vertical fields. In one of them, a slightly higher temperature of the hot gas is found, although 
        the increase is not statistically significant.
\item The lack of a compression-induced heating of the hot gas in the southern outskirts of the disk of NGC\,628 and the simultaneous 
        very high order and low energy density of the magnetic field suggests the shearing of the latter, which together with a distorted spiral arm (S1) 
        could be a sign of tidal interaction. This hypothesis seems to be supported by a sharp gradient of the \ion{H}{i} emission in this area.
\end{itemize}

Although our analyses are not sufficient to provide direct evidence for the role of magnetic reconnection in heating of the ISM, 
the proposed scenario that magnetic reconnection contributes to the heating of the ISM seems feasible. Because this effect is not very strong, 
very sensitive X-ray data are crucial to allow for a statistically significant detection. Studies of edge-on galaxies that show reversals 
of the magnetic field are especially desirable to examine reconnection heating in the galactic halo in more detail.

\begin{acknowledgements}
We thank the anonymous referee for comments that helped to improve the clarity of this paper. 
M.W., M.S., and M.U. are supported by the National Science Centre, Poland, with the grant project 2017/27/B/ST9/01050. 
M. H. acknowledges the support by the National Science Centre through the OPUS grant No. 2015/19/B/ST9/02959.
Research in the field at Ruhr University Bochum is supported by Deutsche Forschungsgemeinschaft SFB 1491.
The National Radio Astronomy Observatory is a facility of the National Science Foundation operated under cooperative agreement
by Associated Universities, Inc.
\end{acknowledgements}

\bibliographystyle{aa} 
\bibliography{myreferences} 

\begin{thebibliography}{49}
\expandafter\ifx\csname natexlab\endcsname\relax\def\natexlab#1{#1}\fi

\bibitem[{{Arnaud}(1996)}]{arnaud96}
{Arnaud}, K.~A. 1996, in Astronomical Society of the Pacific Conference Series,
  Vol. 101, Astronomical Data Analysis Software and Systems V, ed. G.~H.
  {Jacoby} \& J.~{Barnes}, 17

\bibitem[{{Auld} {et~al.}(2006){Auld}, {Minchin}, {Davies}, {Catinella}, {van
  Driel}, {Henning}, {Linder}, {Momjian}, {Muller}, {O'Neil}, {Sabatini},
  {Schneider}, {Bothun}, {Cortese}, {Disney}, {Hoffman}, {Putman}, {Rosenberg},
  {Baes}, {de Blok}, {Boselli}, {Brinks}, {Brosch}, {Irwin}, {Karachentsev},
  {Kilborn}, {Koribalski}, \& {Spekkens}}]{auld06}
{Auld}, R., {Minchin}, R.~F., {Davies}, J.~I., {et~al.} 2006, \mnras, 371, 1617

\bibitem[{{Bagetakos} {et~al.}(2011){Bagetakos}, {Brinks}, {Walter}, {de Blok},
  {Usero}, {Leroy}, {Rich}, \& {Kennicutt}}]{bagetakos11}
{Bagetakos}, I., {Brinks}, E., {Walter}, F., {et~al.} 2011, \aj, 141, 23

\bibitem[{{Beck}(2004)}]{beck04}
{Beck}, R. 2004, \apss, 289, 293

\bibitem[{{Beck}(2007)}]{beck07}
{Beck}, R. 2007, \aap, 470, 539

\bibitem[{{Beck}(2015)}]{beck15}
{Beck}, R. 2015, \aapr, 24, 4

\bibitem[{{Beck} \& {Krause}(2005)}]{beck05}
{Beck}, R. \& {Krause}, M. 2005, Astronomische Nachrichten, 326, 414

\bibitem[{{Carter} \& {Read}(2007)}]{carter07}
{Carter}, J.~A. \& {Read}, A.~M. 2007, \aap, 464, 1155

\bibitem[{{de Avillez}(2000)}]{deavillez00}
{de Avillez}, M.~A. 2000, \mnras, 315, 479

\bibitem[{{De Luca} \& {Molendi}(2004)}]{deluca04}
{De Luca}, A. \& {Molendi}, S. 2004, \aap, 419, 837

\bibitem[{{de Vaucouleurs} {et~al.}(1991){de Vaucouleurs}, {de Vaucouleurs},
  {Corwin}, {Buta}, {Paturel}, \& {Fouque}}]{devaucouleurs91}
{de Vaucouleurs}, G., {de Vaucouleurs}, A., {Corwin}, Herold~G., J., {et~al.}
  1991, {Third Reference Catalogue of Bright Galaxies}

\bibitem[{{Earnshaw} {et~al.}(2019){Earnshaw}, {Roberts}, {Middleton},
  {Walton}, \& {Mateos}}]{earnshaw19}
{Earnshaw}, H.~P., {Roberts}, T.~P., {Middleton}, M.~J., {Walton}, D.~J., \&
  {Mateos}, S. 2019, \mnras, 483, 5554

\bibitem[{{Frick} {et~al.}(2016){Frick}, {Stepanov}, {Beck}, {Sokoloff},
  {Shukurov}, {Ehle}, \& {Lundgren}}]{frick16}
{Frick}, P., {Stepanov}, R., {Beck}, R., {et~al.} 2016, \aap, 585, A21

\bibitem[{{Gabriel} {et~al.}(2004){Gabriel}, {Denby}, {Fyfe}, {Hoar}, {Ibarra},
  {Ojero}, {Osborne}, {Saxton}, {Lammers}, \& {Vacanti}}]{gabriel04}
{Gabriel}, C., {Denby}, M., {Fyfe}, D.~J., {et~al.} 2004, in Astronomical
  Society of the Pacific Conference Series, Vol. 314, Astronomical Data
  Analysis Software and Systems (ADASS) XIII, ed. F.~{Ochsenbein}, M.~G.
  {Allen}, \& D.~{Egret}, 759

\bibitem[{{Hanasz} \& {Lesch}(1998)}]{hanasz98}
{Hanasz}, M. \& {Lesch}, H. 1998, \aap, 332, 77

\bibitem[{{Heald} {et~al.}(2016){Heald}, {de Blok}, {Lucero}, {Carignan},
  {Jarrett}, {Elson}, {Oozeer}, {Randriamampand ry}, \& {van Zee}}]{heald16}
{Heald}, G., {de Blok}, W.~J.~G., {Lucero}, D., {et~al.} 2016, \mnras, 462,
  1238

\bibitem[{{Heintz} {et~al.}(2020){Heintz}, {Bustard}, \& {Zweibel}}]{heintz20}
{Heintz}, E., {Bustard}, C., \& {Zweibel}, E.~G. 2020, \apj, 891, 157

\bibitem[{{HI4PI Collaboration} {et~al.}(2016){HI4PI Collaboration}, {Ben
  Bekhti}, {Fl{\"o}er}, {Keller}, {Kerp}, {Lenz}, {Winkel}, {Bailin},
  {Calabretta}, {Dedes}, {Ford}, {Gibson}, {Haud}, {Janowiecki}, {Kalberla},
  {Lockman}, {McClure-Griffiths}, {Murphy}, {Nakanishi}, {Pisano}, \&
  {Staveley-Smith}}]{hi4pi}
{HI4PI Collaboration}, {Ben Bekhti}, N., {Fl{\"o}er}, L., {et~al.} 2016, \aap,
  594, A116

\bibitem[{{Hu}(2019)}]{hu19}
{Hu}, C.-Y. 2019, \mnras, 483, 3363

\bibitem[{{Jansen} {et~al.}(2001){Jansen}, {Lumb}, {Altieri}, {Clavel}, {Ehle},
  {Erd}, {Gabriel}, {Guainazzi}, {Gondoin}, {Much}, {Munoz}, {Santos},
  {Schartel}, {Texier}, \& {Vacanti}}]{jansen01}
{Jansen}, F., {Lumb}, D., {Altieri}, B., {et~al.} 2001, \aap, 365, L1

\bibitem[{Kaastra(1992)}]{kaastra92}
Kaastra, J.~S. 1992, An X-Ray Spectral Code for Optically Thin Plasmas
  (Internal SRON-Leiden Report, updated version 2.0)

\bibitem[{{Kahn} \& {Brett}(1993)}]{kahn93}
{Kahn}, F.~D. \& {Brett}, L. 1993, \mnras, 263, 37

\bibitem[{{Kamphuis} \& {Briggs}(1992)}]{kamphuis92}
{Kamphuis}, J. \& {Briggs}, F. 1992, \aap, 253, 335

\bibitem[{{Karachentsev} {et~al.}(2004){Karachentsev}, {Karachentseva},
  {Huchtmeier}, \& {Makarov}}]{karachentsev04}
{Karachentsev}, I.~D., {Karachentseva}, V.~E., {Huchtmeier}, W.~K., \&
  {Makarov}, D.~I. 2004, \aj, 127, 2031

\bibitem[{{Kennicutt} {et~al.}(2003){Kennicutt}, {Armus}, {Bendo}, {Calzetti},
  {Dale}, {Draine}, {Engelbracht}, {Gordon}, {Grauer}, {Helou}, {Hollenbach},
  {Jarrett}, {Kewley}, {Leitherer}, {Li}, {Malhotra}, {Regan}, {Rieke},
  {Rieke}, {Roussel}, {Smith}, {Thornley}, \& {Walter}}]{kennicutt03}
{Kennicutt}, Robert~C., J., {Armus}, L., {Bendo}, G., {et~al.} 2003, \pasp,
  115, 928

\bibitem[{{Kennicutt} {et~al.}(2011){Kennicutt}, {Calzetti}, {Aniano},
  {Appleton}, {Armus}, {Beir{\~a}o}, {Bolatto}, {Brandl}, {Crocker}, {Croxall},
  {Dale}, {Donovan Meyer}, {Draine}, {Engelbracht}, {Galametz}, {Gordon},
  {Groves}, {Hao}, {Helou}, {Hinz}, {Hunt}, {Johnson}, {Koda}, {Krause},
  {Leroy}, {Li}, {Meidt}, {Montiel}, {Murphy}, {Rahman}, {Rix}, {Roussel},
  {Sandstrom}, {Sauvage}, {Schinnerer}, {Skibba}, {Smith}, {Srinivasan},
  {Vigroux}, {Walter}, {Wilson}, {Wolfire}, \& {Zibetti}}]{kennicutt11}
{Kennicutt}, R.~C., {Calzetti}, D., {Aniano}, G., {et~al.} 2011, \pasp, 123,
  1347

\bibitem[{{Lazarian} \& {Vishniac}(1999)}]{lazarian99}
{Lazarian}, A. \& {Vishniac}, E.~T. 1999, \apj, 517, 700

\bibitem[{{Lin} {et~al.}(2012){Lin}, {Webb}, \& {Barret}}]{lin12}
{Lin}, D., {Webb}, N.~A., \& {Barret}, D. 2012, \apj, 756, 27

\bibitem[{{Lundgren} {et~al.}(2004){Lundgren}, {Wiklind}, {Olofsson}, \&
  {Rydbeck}}]{lundgren04}
{Lundgren}, A.~A., {Wiklind}, T., {Olofsson}, H., \& {Rydbeck}, G. 2004, \aap,
  413, 505

\bibitem[{{Makarov} {et~al.}(2014){Makarov}, {Prugniel}, {Terekhova},
  {Courtois}, \& {Vauglin}}]{makarov14}
{Makarov}, D., {Prugniel}, P., {Terekhova}, N., {Courtois}, H., \& {Vauglin},
  I. 2014, \aap, 570, A13

\bibitem[{{Marcum} {et~al.}(2001){Marcum}, {O'Connell}, {Fanelli}, {Cornett},
  {Waller}, {Bohlin}, {Neff}, {Roberts}, {Smith}, {Cheng}, {Collins},
  {Hennessy}, {Hill}, {Hill}, {Hintzen}, {Landsman}, {Ohl}, {Parise}, {Smith},
  {Freedman}, {Kuchinski}, {Madore}, {Angione}, {Palma}, {Talbert}, \&
  {Stecher}}]{marcum01}
{Marcum}, P.~M., {O'Connell}, R.~W., {Fanelli}, M.~N., {et~al.} 2001, \apjs,
  132, 129

\bibitem[{{Mewe} {et~al.}(1985){Mewe}, {Gronenschild}, \& {van den
  Oord}}]{mewe85}
{Mewe}, R., {Gronenschild}, E.~H.~B.~M., \& {van den Oord}, G.~H.~J. 1985,
  \aaps, 62, 197

\bibitem[{{Mora-Partiarroyo} {et~al.}(2019){Mora-Partiarroyo}, {Krause},
  {Basu}, {Beck}, {Wiegert}, {Irwin}, {Henriksen}, {Stein}, {Vargas}, {Heesen},
  {Walterbos}, {Rand}, {Heald}, {Li}, {Kamieneski}, \& {English}}]{mora19}
{Mora-Partiarroyo}, S.~C., {Krause}, M., {Basu}, A., {et~al.} 2019, \aap, 632,
  A11

\bibitem[{{Mulcahy} {et~al.}(2017){Mulcahy}, {Beck}, \& {Heald}}]{mulcahy17}
{Mulcahy}, D.~D., {Beck}, R., \& {Heald}, G.~H. 2017, \aap, 600, A6

\bibitem[{{Nulsen} {et~al.}(1984){Nulsen}, {Stewart}, \& {Fabian}}]{nulsen84}
{Nulsen}, P.~E.~J., {Stewart}, G.~C., \& {Fabian}, A.~C. 1984, \mnras, 208, 185

\bibitem[{{Owen} \& {Warwick}(2009)}]{owen09}
{Owen}, R.~A. \& {Warwick}, R.~S. 2009, \mnras, 394, 1741

\bibitem[{{Raymond} {et~al.}(1976){Raymond}, {Cox}, \& {Smith}}]{raymond76}
{Raymond}, J.~C., {Cox}, D.~P., \& {Smith}, B.~W. 1976, \apj, 204, 290

\bibitem[{{Sonba{\c{s}}} {et~al.}(2010){Sonba{\c{s}}}, {Aky{\"u}z}, {Balman},
  \& {{\"O}zel}}]{sonbas10}
{Sonba{\c{s}}}, E., {Aky{\"u}z}, A., {Balman}, {\c{S}}., \& {{\"O}zel}, M.~E.
  2010, \aap, 517, A91

\bibitem[{{Str{\"u}der} {et~al.}(2001){Str{\"u}der}, {Briel}, {Dennerl},
  {Hartmann}, {Kendziorra}, {Meidinger}, {Pfeffermann}, {Reppin}, {Aschenbach},
  {Bornemann}, {Br{\"a}uninger}, {Burkert}, {Elender}, {Freyberg}, {Haberl},
  {Hartner}, {Heuschmann}, {Hippmann}, {Kastelic}, {Kemmer}, {Kettenring},
  {Kink}, {Krause}, {M{\"u}ller}, {Oppitz}, {Pietsch}, {Popp}, {Predehl},
  {Read}, {Stephan}, {St{\"o}tter}, {Tr{\"u}mper}, {Holl}, {Kemmer}, {Soltau},
  {St{\"o}tter}, {Weber}, {Weichert}, {von Zanthier}, {Carathanassis}, {Lutz},
  {Richter}, {Solc}, {B{\"o}ttcher}, {Kuster}, {Staubert}, {Abbey}, {Holland},
  {Turner}, {Balasini}, {Bignami}, {La Palombara}, {Villa}, {Buttler},
  {Gianini}, {Lain{\'e}}, {Lumb}, \& {Dhez}}]{strueder01}
{Str{\"u}der}, L., {Briel}, U., {Dennerl}, K., {et~al.} 2001, \aap, 365, L18

\bibitem[{{Tanuma} {et~al.}(2003){Tanuma}, {Yokoyama}, {Kudoh}, \&
  {Shibata}}]{tanuma03}
{Tanuma}, S., {Yokoyama}, T., {Kudoh}, T., \& {Shibata}, K. 2003, \apj, 582,
  215

\bibitem[{{T{\"u}llmann} {et~al.}(2006){T{\"u}llmann}, {Pietsch}, {Rossa},
  {Breitschwerdt}, \& {Dettmar}}]{tuellmann06}
{T{\"u}llmann}, R., {Pietsch}, W., {Rossa}, J., {Breitschwerdt}, D., \&
  {Dettmar}, R.-J. 2006, \aap, 448, 43

\bibitem[{{Turner} {et~al.}(2001){Turner}, {Abbey}, {Arnaud}, {Balasini},
  {Barbera}, {Belsole}, {Bennie}, {Bernard}, {Bignami}, {Boer}, {Briel},
  {Butler}, {Cara}, {Chabaud}, {Cole}, {Collura}, {Conte}, {Cros}, {Denby},
  {Dhez}, {Di Coco}, {Dowson}, {Ferrando}, {Ghizzardi}, {Gianotti}, {Goodall},
  {Gretton}, {Griffiths}, {Hainaut}, {Hochedez}, {Holland}, {Jourdain},
  {Kendziorra}, {Lagostina}, {Laine}, {La Palombara}, {Lortholary}, {Lumb},
  {Marty}, {Molendi}, {Pigot}, {Poindron}, {Pounds}, {Reeves}, {Reppin},
  {Rothenflug}, {Salvetat}, {Sauvageot}, {Schmitt}, {Sembay}, {Short},
  {Spragg}, {Stephen}, {Str{\"u}der}, {Tiengo}, {Trifoglio}, {Tr{\"u}mper},
  {Vercellone}, {Vigroux}, {Villa}, {Ward}, {Whitehead}, \& {Zonca}}]{turner01}
{Turner}, M.~J.~L., {Abbey}, A., {Arnaud}, M., {et~al.} 2001, \aap, 365, L27

\bibitem[{{Vishniac} \& {Lazarian}(1999)}]{vishniac99}
{Vishniac}, E.~T. \& {Lazarian}, A. 1999, \apj, 511, 193

\bibitem[{{Walter} {et~al.}(2008){Walter}, {Brinks}, {de Blok}, {Bigiel},
  {Kennicutt}, {Thornley}, \& {Leroy}}]{walter08}
{Walter}, F., {Brinks}, E., {de Blok}, W.~J.~G., {et~al.} 2008, \aj, 136, 2563

\bibitem[{{Walton} {et~al.}(2011){Walton}, {Gladstone}, {Roberts}, \& {Fabian
  }}]{walton11}
{Walton}, D.~J., {Gladstone}, J.~C., {Roberts}, T.~P., \& {Fabian }, A.~C.
  2011, Astronomische Nachrichten, 332, 354

\bibitem[{{We{\.z}gowiec} {et~al.}(2012){We{\.z}gowiec}, {Bomans}, {Ehle},
  {Chy{\.z}y}, {Urbanik}, {Braine}, \& {Soida}}]{wezgowiec12}
{We{\.z}gowiec}, M., {Bomans}, D.~J., {Ehle}, M., {et~al.} 2012, \aap, 544, A99

\bibitem[{{We{\.z}gowiec} {et~al.}(2016){We{\.z}gowiec}, {Ehle}, \&
  {Beck}}]{wezgowiec16}
{We{\.z}gowiec}, M., {Ehle}, M., \& {Beck}, R. 2016, \aap, 585, A3

\bibitem[{{We{\.z}gowiec} {et~al.}(2020){We{\.z}gowiec}, {Ehle}, {Soida},
  {Dettmar}, {Beck}, \& {Urbanik}}]{wezgowiec20}
{We{\.z}gowiec}, M., {Ehle}, M., {Soida}, M., {et~al.} 2020, \aap, 640, A109

\bibitem[{{Zimmer} {et~al.}(1996){Zimmer}, {Birk}, {Kerp}, \&
  {Lesch}}]{zimmer96}
{Zimmer}, F., {Birk}, G., {Kerp}, J., \& {Lesch}, H. 1996, Astrophysical
  Letters and Communications, 34, 193

\end{thebibliography}

\end{document}